\pdfoutput=1
\documentclass[fleqn,usenatbib,usedcolumn]{mnras}

\usepackage{graphicx}	
\usepackage{amsmath}	
\usepackage{amssymb}	
\usepackage{csquotes}

\usepackage{txfonts}

\usepackage[T1]{fontenc}
\usepackage{ae,aecompl}

\usepackage{multirow}
\usepackage{caption}
\usepackage{dsfont}
\usepackage[capitalise]{cleveref}
\usepackage{color}
\usepackage{arydshln}
\usepackage{natbib}

\hypersetup{pdfauthor={J. Salcido},
               pdftitle={Music from the heavens - Gravitational waves from supermassive black hole mergers in the EAGLE simulations},
               pdfkeywords={black hole physics, gravitational waves, cosmology: theory, early Universe, galaxies: formation, galaxies: evolution},
               bookmarksnumbered=true}

\volume{{\rm in press}}

\usepackage{etoolbox}
\makeatletter
\patchcmd\@combinedblfloats{\box\@outputbox}{\unvbox\@outputbox}{}{%
  \errmessage{\noexpand\@combinedblfloats could not be patched}%
}%
\makeatother

\newcommand{\noop}[1]{}

\title[GWs from SMBH mergers in the EAGLE simulations]{Music from the heavens - Gravitational waves from supermassive black hole mergers in the EAGLE simulations}

\author[J. Salcido et al.]{Jaime Salcido,$^{1}$\thanks{E-mail:
\href{mailto:jaime.salcido@durham.ac.uk}{jaime.salcido@durham.ac.uk}} Richard G. Bower,$^{1}$ Tom Theuns,$^{1}$ Stuart McAlpine,$^{1}$
\newauthor Matthieu Schaller,$^{1}$ Robert A. Crain,$^{2}$ Joop Schaye,$^{3}$ John Regan$^{1}$
\\
$^{1}$ Institute for Computational Cosmology, Department of Physics, Durham University, South Road, Durham, DH1 3LE, UK\\
$^{2}$ Astrophysics Research Institute, Liverpool John Moores University, 146 Brownlow Hill, Liverpool L3 5RF, UK\\
$^{3}$ Leiden Observatory, Leiden University, P.O. Box 9513, 2300 RA Leiden, the Netherlands}

\date{Accepted XXX. Received YYY; in original form ZZZ}

\pubyear{2015}

\begin{document}
\label{firstpage}
\pagerange{\pageref{firstpage}--\pageref{lastpage}}
\maketitle

\begin{abstract}

We estimate the expected event rate of gravitational wave signals from mergers of supermassive black holes that could be resolved by a space-based interferometer, such as the Evolved Laser Interferometer Space Antenna (eLISA), utilising the reference cosmological hydrodynamical simulation from the \textsc{eagle} suite. These simulations assume a $\Lambda$CDM cosmogony with state-of-the-art subgrid models for radiative cooling, star formation, stellar mass loss, and feedback from stars and accreting black holes. They have been shown to reproduce the observed galaxy population with unprecedented fidelity. We combine the merger rates of supermassive black holes in \textsc{eagle} with the latest phenomenological waveform models to calculate the gravitational waves signals from the intrinsic parameters of the merging black holes. The \textsc{eagle} models predict $\sim2$ detections per year by a gravitational wave detector such as eLISA. We find that these signals are largely dominated by mergers between seed mass black holes merging at redshifts between $z\sim2$ and $z\sim1$. In order to investigate the dependence on the assumed black hole seed mass, we introduce an additional model with a black hole seed mass an order of magnitude smaller than in our reference model. We also consider a variation of the reference model where a prescription for the expected delays in the black hole merger timescale has been included after their host galaxies merge. We find that the merger rate is similar in all models, but that the initial black hole seed mass could be distinguished through their detected gravitational waveforms. Hence, the characteristic gravitational wave signals detected by eLISA will provide profound insight into the origin of supermassive black holes and the initial mass distribution of black hole seeds.  
\end{abstract}




\section{Introduction}
In our current understanding of extragalactic astrophysics supermassive black holes (SMBHs) reside at the centres of most galaxies at $z=0$ and were responsible for powering the luminous quasars observed within the first billion years of the Universe (e.g. \citealt{Fan2006,VolonteriBellovary:2012}). Observations of a tight correlation between the mass of a galaxy's central SMBH and key properties of its galactic host, such as the bulge mass and stellar velocity dispersion (e.g. \citealt{Magorrian:1998,Gebhardt:2000,Ferrarese:2000,Gebhardt:2009}), have led to the idea that SMBHs  play a major role in the evolution of their host galaxies (e.g. \citealt{Bower:2006,Volonteri-Bellovay:2011-evol,Fabian:2012,Alexander:2012,Kormendy:2013}). It seems, therefore, that feedback from active galactic nuclei (AGN), galaxy mergers, and the growth of SMBHs are closely intertwined (e.g. \citealt{Kauffmann:2000,King:2003,DiMatteo:2005,Booth:2009,Fanidakis:2011}).
 
In a standard Lambda Cold Dark Matter ($\Lambda$CDM) cosmology cosmic structures build up hierarchically by the continuous merging of smaller structures and the accretion of surrounding matter. In this hierarchical scenario central SMBHs follow a similar build-up process and are the result of a complex evolution, in which black hole (BH) seeds grow both through accretion episodes and mergers with other BHs. However, constraining the formation mechanisms of BHs represents a major observational challenge. The direct detection of gravitational wave (GW) signals from SMBH mergers may prove to be a viable way to discriminate among the different BH seed formation models \citep[e.g.][]{Volonteri:2010,Seoane:2012eLISA}. The discovery of the GW source GW150914 by the LIGO collaboration provided the first observational evidence for the existence of binary BH systems that inspiral and merge within the Hubble time \citep{LIGO}. The gravitational radiation emitted during the merging of SMBHs in the centres of colliding galaxies will produce some of the \enquote*{\textit{loudest}} events in the Universe, which can provide us with unique information about the nature of BHs and also provides a test of our understanding of gravity and galaxy evolution.

In the last decade major efforts have been made to predict the event rate of GWs in the frequency band of a space-based GW detector such as the Evolved Laser Interferometer Space Antenna (eLISA,  \citealt{Seoane:2012eLISA, Seoane:2012GWN}). These predictions range from a few, up to a few hundred events per year, depending on the assumptions underpinning the calculation of the SMBHs coalescence rate. Early works derived the SMBH coalescence rate from observational constraints such as the observed quasar luminosity function \citep{Haehnelt:1994}, whilst more recent studies have utilised semi-analytical galaxy formation models and/or hybrid models that combine cosmological N-body simulations with semi-analytical recipes for the SMBH dynamics \citep[e.g.][]{WyitheLoeb:2003,Enoki:2004,Koushiappas:2005,Sesana:2007,Micic:2007,Sesana:2009-SA,Klein:2016}. In contrast to semi-analytic models, hydrodynamical simulations follow the dynamics of the cosmic gas by direct numerical integration of the equations of hydrodynamics capturing non-linear processes that cannot be described by simple mathematical approximations. Hence a more complete and consistent picture of the evolution of SMBHs and their host galaxies can be obtained.

The \textit{Evolution and Assembly of GaLaxies and their Environment} (\textsc{eagle}) project \citep{Schaye:2015,Crain:2015} consists of a suite of hydrodynamical simulations of a $\Lambda$CDM cosmogony. Using state-of-the-art subgrid models for radiative cooling, star formation, stellar mass loss, and feedback from stars and accreting BHs, the simulations reproduce the observed galaxy population with unprecedented fidelity. Key observations, such as the present-day stellar mass function of galaxies, the dependence of galaxy sizes on stellar mass, and the amplitude of the central BH mass-stellar mass relation, as well as many other properties of observed galaxies and the intergalactic medium (both at the present day and at earlier epochs) are reproduced by the simulations \citep[e.g.][]{Furlong:2015-Sizes,Furlong:2015,Trayford:2015,Schaller:2015,Lagos:2015,Rahmati:2015,Rahmati:2015b, Bahe:2016, Rosas-Guevara:2016}. In this study we introduce the first estimate of the event rate of GWs expected from SMBH mergers utilising large-scale cosmological hydrodynamical sim ulations. We compute that the event rate of GW signals is low enough to produce a set of events that are resolvable by a space-based interferometer, such as eLISA. 

The layout of this paper is as follows: In \cref{sec:GW} we provide a brief summary of the basic equations of the GW signals produced by the SMBH coalescence process. \Cref{sec:Sim} presents a brief overview of the \textsc{eagle} simulation suite, including the list of simulations used in this study, a discussion of the BH seeding mechanism and growth, as well as the calculated SMBH merger rates from the simulations. In \cref{sec:results} we present the predicted GW signals from the simulations and discuss our main results. We discuss the limitations of our analysis, making some remarks on the simulations and the SMBH seeding model adopted in \textsc{eagle} and conclude in \cref{sec:con}. 

\begin{figure}\centering
 \includegraphics[width=0.48\textwidth]{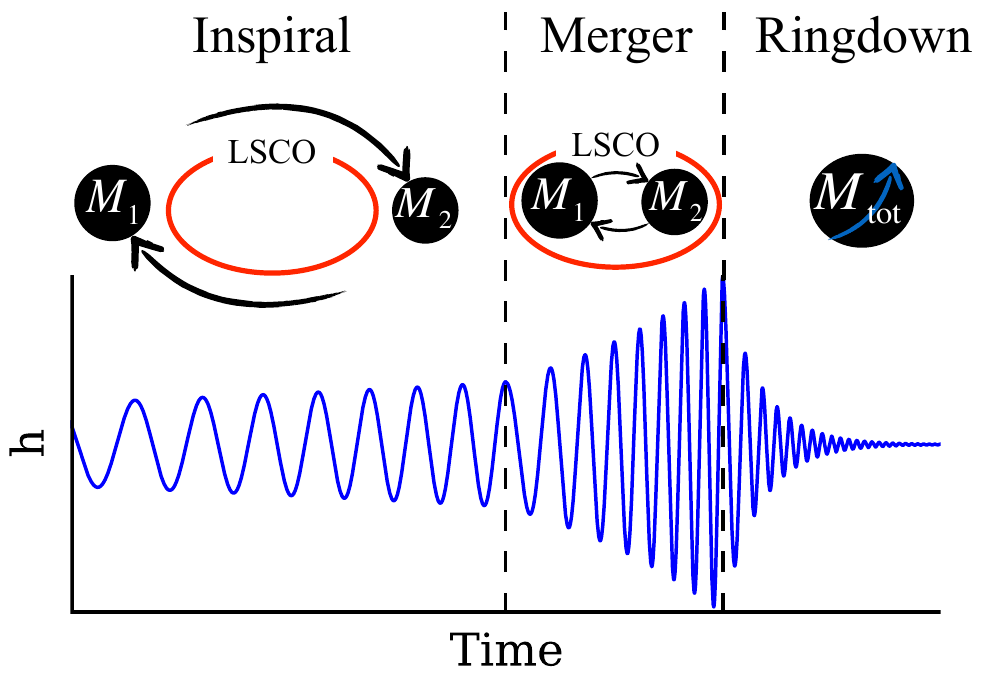}
 \vspace{-1.5em}
 \caption{Schematic diagram of the phase evolution (inspiral, merger, and ringdown) of a non-spinning SMBH binary coalescence process. The Last Stable Circular Orbit (LSCO) of the binary is shown as the red curve. The resulting SMBH may be rapidly rotating even if the progenitor BHs had very small or no spin \citep{Flanagan:1997}. Below each phase an example of the \textit{strain}   amplitude, $h$, as a function of time is shown for the dominant spherical harmonic mode of the GW signal from the non spinning SMBH binary. This specifies the fractional change in the relative displacement between freely falling test masses in a detector due to the GW.}
 \label{fig:waveforms}
\end{figure}
 
\begin{figure*}
 \includegraphics[width=0.99\textwidth]{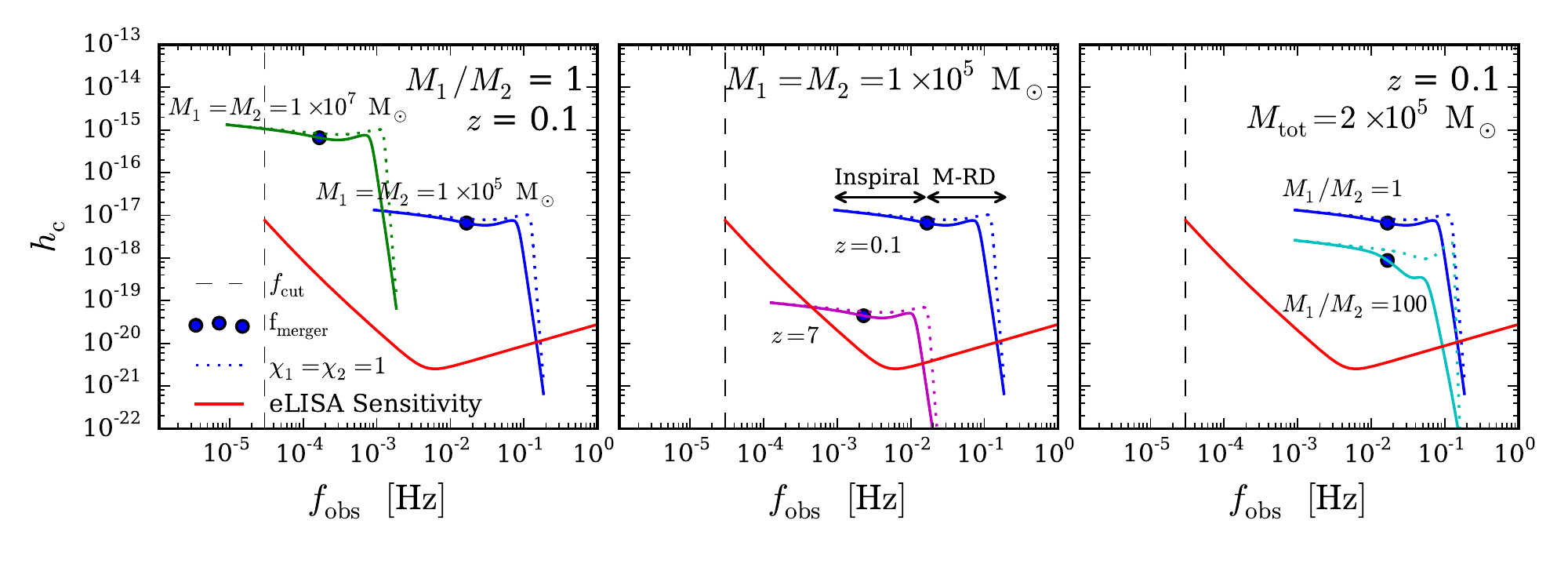}
 \vspace{-1.5em}
 \caption{Example of the dimensionless characteristic strain amplitude, $h_c$, produced by non-spinning SMBH coalescence events ($\chi_i = 0$). Maximally spinning SMBHs aligned with the orbital angular momentum of the binary are shown in dotted lines ($\chi_i = 1$). In all panels the inspiral and merger-ringdown phases are shown for an equal mass BH binary (Mass ratio ${M}_1/{M}_2=1$, ${M}_1 = {M}_2 = 1 \times 10^5 \,\, \mathrm{M}_{\odot}$) that merge at redshift $z=0.1$ as reference (blue line). The frequency at the transition from the inspiral phase to the merger phase ($f_{\mathrm{merger}} = 0.018 \, c^3/GM_{\mathrm{total}}$) is highlighted with a blue dot. The sensitivity curve of eLISA was calculated from the analytic approximation provided by \citet{Seoane:2012GWN}. The black dashed line indicates the low-frequency cut-off of the sensitivity curve $f_\mathrm{cut} = 3 \times 10^{-5} \,\, \mathrm{Hz}$. GW signals above the sensitivity curve and to the right of the low-frequency cut-off can be resolved from the eLISA data stream. \textit{LEFT PANEL:} The effect of increasing the total mass of the SMBH binary. An equal mass SMBH binary (Mass ratio ${M}_1/{M}_2=1$) with ${M}_1 = {M}_2 = 1 \times 10^7 \,\, \mathrm{M}_{\odot}$ that merges at redshift $z=0.1$ is shown in green. \textit{MIDDLE PANEL}: The effect of redshift. An equal mass SMBH binary (Mass ratio ${M}_1/{M}_2=1$, ${M}_1 = {M}_2 = 1 \times 10^5 \,\, \mathrm{M}_{\odot}$) merging at redshift $z=7$ is shown in magenta. \textit{RIGHT  PANEL}: The effect of mass ratio. A BH binary with total mass ${M}_{\mathrm{total}} = 2 \times 10^5 \,\, \mathrm{M}_{\odot}$ and mass ratio ${M}_1/{M}_2=100$ merging at redshift $z=0.1$ is shown in cyan.}
 \label{fig:spectrum}
\end{figure*}

\section{Gravitational Wave emission from SMBH mergers}\label{sec:GW}
 
Studying SMBH mergers involve physical processes that cover many orders of magnitude in physical size. From mergers of dark matter haloes and galaxies driven by the cosmic web at large scales (>Mpc), to the final SMBH merger via the emission of GWs that occurs at sub-parsec scales. The overall scenario was first outlined by \cite*{Begelman:1980}. When two dark matter halos merge, the galaxies that they host will eventually merge by the effect of dynamical friction. After the galaxy merger, the central SMBHs are brought near the centre of the main halo due to dynamical friction against the dark matter, background stars, and gas. The efficiency of dynamical friction decreases when the SMBHs become close and form a bound binary \citep{Mayer:2007}. The dynamical evolution of the SMBH binary is expected to be radically different in gas-rich and gas-poor galaxies. In gas-rich galaxies planet-like migration can effectively dissipate energy and angular momentum from the binary, leading to a short coalescence time-scale, typically in the order of $\sim10^7 - 10^8\, \mathrm{yrs}$ (\citealt{Escala:2005,Colpi:2014} but see e.g. \citealt{Tamburello:2016}). In gas-poor systems, the evolution of the binary is largely determined by three-body interactions with the background stars, leading to a long coalescence timescale of the order of a few Gyrs. At milliparsec separations, GW emission drives the final coalescence process (see \citealt{Colpi:2011}, \citealt{Mayer:2013} and \citealt{Colpi:2014} for a review on SMBH dynamics in galaxy mergers).

The BH binary merging process can be divided in three distinct phases which are illustrated by \cref{fig:waveforms}:
\begin{enumerate}
	\item The \textbf{inspiral phase}, during which the distance between the bound SMBH binary is larger than the Last Stable Circular Orbit ($R_\mathrm{LSCO}$) and the mutual gravitational field strength is weak. Since the location of the LSCO is very difficult to calculate for a binary BH system, here $R_\mathrm{LSCO}$ is approximated by the limiting case of a test particle orbiting a non-rotating BH, $R_\mathrm{LSCO} = 6GM_1/c^2 = 3R_S$, where $R_S$ is the Schwarzschild radius of the most massive BH in the binary. Post-Newtonian equations provide an accurate representation of the dynamical evolution of the binary in this phase. The GW signals emitted during the inspiral phase have a characteristic shape with slowly increasing frequency and amplitude. \\
	\item In the highly non-linear \textbf{merger phase} SMBHs approach to within the last stable circular orbit ($ \leq R_\mathrm{LSCO}$) and the dynamics evolve to a rapid plunge and coalescence. In this regime the event horizons of the BHs overlap and the geometry of the local spacetime becomes extremely complicated. Analytical schemes break down in this regime and numerical relativity (NR) is needed to model the dynamics through the merger phase. \\
	\item Finally, the \textbf{quasi-normal-ringdown phase}, where the resulting BH settles down to a rotating Kerr BH emitting GWs due to its deviations from the final axisymmetric state. Perturbation theory can be applied to obtain the quasi-normal modes in this phase. The GW signal emitted during the ringdown phase has a characteristic shape consisting of the superposition of exponentially damped sinusoids.
	\end{enumerate}

In general relativity the \textit{`no-hair'} theorem posits that BHs are entirely characterised by only three parameters, namely their mass, spin, and electric charge. For astrophysical BHs the electric charge is usually expected to be negligible \citep[pp. 875-876]{Gravitation:1973}. Therefore each coalescing SMBH is fully characterised by the total mass $M_\mathrm{total} = M_1+M_2$, the mass ratio $M_1/M_2$ of the binary and the BH spin angular momenta $\vec{S}_1$ and $\vec{S}_2$. ${M}_1$ is defined as the more massive member of the BH binary (${M}_1 \geq {M}_2$). The most general detectable signal from a SMBH binary is a function of the  intrinsic properties of the binary, the merger redshift $z$, and the observer's orientation. In this study for simplicity we focus on non-spinning SMBH binaries as potential sources of GWs. In \cref{sec:App1} we extend our analysis to investigate the case of rapidly spinning coalescing SMBHs. The inclusion of the signal from the ringdown phase increases the signal-to-noise ratio (S/N) of observed binaries and enable measurements of the parameters of the resulting SMBH \citep{Klein:2016}.

\subsection{Characteristic strain}\label{sec:h_c}

It is difficult to determine accurately the total gravitational energy radiated as GWs in a BH binary coalescence and modelling the GW signal from these processes still represents a challenge for GW astronomy \citep{Ohme:2012,Hannam2014}.

In general, the total energy radiated during a BH coalescence will be some fraction, $\epsilon$, of the total rest mass energy of the binary (${M}_{\mathrm{total}}c^2$) that depends on the mass ratio ${M}_1/{M}_2$, the orbital angular momentum and the initial spin of the BHs. 

In GW astronomy it is common to describe the amplitude of a source using the dimensionless \textit{strain} as a function of time, $h(t)$. This specifies the fractional change in the relative displacement between two test masses, $h(t)=\Delta L(t)/L_0$, where $L$ is the distance between free-falling masses that constitute the GW detector. The dimensionless characteristic strain amplitude, $h_c$, is not the instantaneous strain of a source but rather an accumulated signal, intended to include the effect of integrating a signal during the inspiralling phase. The characteristic strain amplitude, $h_c$, is defined as
\begin{equation}
	\left[h_c(f_s)\right]^2=4f^2\left|\tilde{h}(f_s)\right|^2,
\end{equation}
where $\tilde{h}(f_s)$ is the Fourier transform of the strain signal, ${\tilde{h}(f_s)=\mathcal{F}\{h(t)\}(f_s)=\int^\infty_{-\infty} h(t)e^{-2\pi if_st} \mathrm{d}t}$, and $f_s = f_{\mathrm{obs}}(1+z)$ is the rest-frame frequency of the signal \citep{Moore:2015}.

We employ the most recent phenomenological frequency-domain gravitational waveform model for non-precessing BH binaries described in \citet{PhenomD2} (commonly referred to as ``PhenomD''). The PhenomD model provides the waveform families in the Fourier domain of the dominant spherical-harmonic modes of the GW signal in aligned-spin systems in terms of the signal amplitude, $A(f_s)$, and phase, $\phi(f_s)$, given by
\begin{equation}
	\tilde{h}(f_s) = A(f_s) e^{-i\phi(f_s)}.
\end{equation}
In this hybrid model, the inspiral and merger-ringdown parts of the signal are modelled separately in two frequency regimes of the waveform. The first region covers the inspiral signal, up to the merger frequency
\begin{equation}\label{eq:fmerger}
f_{\mathrm{merger}} = 0.018 \, c^3/GM_{\mathrm{total}},
\end{equation}
which is approximately the frequency at the LSCO of a test particle orbiting a non-rotating BH ($R_\mathrm{LSCO}=6GM_1/c^2$) \citep{Flanagan:1997}. In the inspiral region, analytic post-Newtonian prescriptions and effective-one-body methods are used to describe the signal. 

The second region (which is  is further sub-divided into intermediate and merger-ringdown regions) uses phenomenological models calibrated to pure NR simulations to describe the signal. The full waveform strain signal is parameterised by the physical parameters of the BH binary, total mass ($M_{\mathrm{total}}=M_1+M_2$), luminosity distance ($D_\mathrm{L}(z)$), symmetric mass ratio ($\eta = M_1M_2/M_{\mathrm{total}}^2$), and the dimensionless spin parameters defined as
\begin{equation}
	\chi_i = \frac{\vec{S}_i \cdot \hat{L}} {M_i^2},
\end{equation}
where $\chi_i \in [-1,1]$ and the BH spin angular momenta, $\vec{S}_i$, are assumed to be parallel to the direction of the orbital angular momentum, $\hat{L}$. 

The loss of energy through GWs leads to a decrease in the separation of the BH binary and hence the orbital frequency increases. For Keplerian circular orbits the frequency of the GWs is twice the orbital frequency $(f_s = 2 f_{\mathrm{orbit}})$. Integrating the frequency evolution of the inspiral phase, or chirp, $\dot{f} = \mathrm{d} f/\mathrm{d} t$, we can estimate the time it takes for the binary to evolve between any two frequencies. It can be shown \citep{Shapiro:1983,Tinto:1988} that for BH binaries on Keplerian circular orbits
\begin{equation}\label{eq:t_ins}
\begin{aligned}
	t(f_2) - t(f_1)  = & \frac{5}{256}\frac{c^5}{G^{5/3}}  \frac{\left( M_1 + M_2 \right)^{1/3}}{M_1M_2} \\ & \times (2\pi)^{-8/3}\left(f_1^{-8/3} - f_2^{-8/3} \right).
\end{aligned}
\end{equation}

The intrinsic duration of the inspiral phase is then given by
\begin{equation}\label{eq:t_insTE}
\begin{aligned}
	\tau_\mathrm{inspiral} & = t(f_\mathrm{merger}) - t(f_\mathrm{min}), 
	\end{aligned}
\end{equation}
where the value of $f_{\mathrm{min}}$ at which the inspiral spectrum starts is uncertain. Since $f_\mathrm{merger}$ is set by \cref{eq:fmerger}, $f_{\mathrm{min}}$ is a free parameter in our calculations. Clearly, choosing $f_{\mathrm{min}}$ very close to $f_\mathrm{merger}$ gives short inspiral times. On the other hand, $f_{\mathrm{min}} \ll f_\mathrm{merger}$ will produce the opposite effect. Following the approach of \citet{Koushiappas:2005}, we choose $f_{\mathrm{min}}$ to be close to $f_{\mathrm{merger}}$ to ensure reasonable values for the time that the BH binary systems spend in the inspiral phase, which are also comparable to the orbital frequency at the BH binary hardening radius \citep{Quinlan:1996}. We have chosen $f_{\mathrm{min}} = 1\times 10^{-3} \, c^3/GM_{\mathrm{total}}$  for our analysis.  

The intrinsic duration of the merger phase is approximated by \citep{Koushiappas:2005},
\begin{equation}\label{eq:t_mer}
	\tau_\mathrm{merger}\sim 14.7 \left(\frac{M_\mathrm{total}}{1 \times 10^{5} \,\mathrm{M}_{\sun}}\right) \, \left[\mathrm{sec}\right].
\end{equation}

\citet{PhenomD2} show that the phenomenological approach is capable of describing waveforms from BH binaries with a high degree of physical fidelity. The range of calibration of the model is: mass-ratio $\in [1, 18]$ and spins $\in [-0.95, 0.98]$. Nonetheless, the model can be evaluated at any physically allowed mass-ratio. In our study, it was required to extend the use of the PhenomD model outside its calibration region for some SMBH binary cases, which can produce physically plausible results that are reasonable for our event rate estimations. However, individual binary parameter estimation would require one to check the model in this region against fully general relativistic NR calculations. More details about the PhenomD model can be found in \citet{PhenomD2}.
 
In \cref{fig:spectrum} we show some examples of the dimensionless characteristic strain amplitude $h_c$ produced by massive BH coalescence events with different masses and occurring at different redshifts, calculated with the equations given in this section. 

\subsection{The eLISA sensitivity curve}

The Evolved Laser Interferometer Space Antenna is a space-based mission designed to measure gravitational radiation over a broad band of frequencies ranging between $f \sim 0.1 \,\, \mathrm{mHz}$ to $f \sim 1 \,\, \mathrm{Hz}$. The final design specification of the mission are yet to be evaluated, including key features like the low-frequency acceleration noise, mission lifetime, the length of the interferometer arms, and the number of laser links between the spacecraft \citep{Klein:2016}. In our study we will use the New Gravitational Observatory (NGO) concept, which was proposed to the European Space Agency (ESA) during the selection process for the L1 large satellite mission. We refer the reader to \citet{Seoane:2012GWN} for a detailed description of the eLISA concept and architecture. 

According to the design requirements, the sensitivity that eLISA will be able to achieve in dimensionless characteristic strain noise amplitude is
\begin{equation}\label{eq:sensitivity}
	h_n(f_{\mathrm{obs}})=\sqrt{S_n(f_{\mathrm{obs}})f_{\mathrm{obs}}},
\end{equation}
where the strain noise power spectral density $S_n(f_{\mathrm{obs}})$ is given by the analytical approximation,
\begin{equation}
\begin{aligned}
	S_n(f_{\mathrm{obs}}) = & \frac{20}{3} \frac{4 \times S_{\mathrm{acc}}(f_{\mathrm{obs}}) + S_{\mathrm{sn}}(f_{\mathrm{obs}}) + S_{\mathrm{omn}}(f_{\mathrm{obs}})}{L^2} \\
				 & \times \left(1+\left(\frac{f_{\mathrm{obs}}}{0.41\left(\frac{c}{2L}\right)}\right)^2\right),
\end{aligned}
\end{equation}
where $f_{\mathrm{obs}}$ is the observed frequency and $L=1\times10^9 \,\, \mathrm{m}$ is the optical path-length between the free-falling masses. 
At low frequencies the noise spectrum of the instrument is dominated by residual acceleration noise of the test masses caused by force gradients arising due to the relative movement of the spacecraft with respect to the test masses
\begin{equation}
	S_{\mathrm{acc}}(f_{\mathrm{obs}}) = 1.37 \times 10^{-32} \left(1+\frac{10^{-4}\mathrm{Hz}}{f_{\mathrm{obs}}}\right)\left(\frac{\mathrm{Hz}}{f_{\mathrm{obs}}}\right)^4 \,\, \left[\mathrm{m}^2\mathrm{Hz}^{-1}\right].
\end{equation}
For $f_\mathrm{obs}\gtrsim 5\times10^{-3} \, \mathrm{Hz}$, the arm length measurement noise dominates, for which the quantum mechanical photon shot noise is
\begin{equation}
\begin{aligned}
	S_{\mathrm{sn}}(f_{\mathrm{obs}}) = 5.25 \times 10 ^{-23} \,\, \left[\mathrm{m}^2\mathrm{Hz}^{-1}\right].
\end{aligned}
\end{equation}
At higher frequencies the sensitivity decreases again, due to the arm-length response to multiple wavelengths of  GWs. This effect is included with other combined measurement noise in,
\begin{equation}\label{eq:noise3}
\begin{aligned}
	S_{\mathrm{omn}}(f_{\mathrm{obs}}) = 6.28 \times 10 ^{-23} \,\, \left[\mathrm{m}^2\mathrm{Hz}^{-1}\right].
\end{aligned}
\end{equation}
The eLISA sensitivity curve obtained from \cref{eq:sensitivity} is plotted as the red curves in \cref{fig:spectrum}. 

The measurement frequency bandwidth requirement for the detector is ($10^{-4}\,\mathrm{Hz}$ to $1\,\mathrm{Hz}$) with a target of ($3\times10^{-5}\,\mathrm{Hz}$ to $1\,\mathrm{Hz}$) \citep{Seoane:2012GWN}. For our analysis we adopt the target frequency cut $f_{\mathrm{cut}} = 3 \times 10^{-5} \,\, \mathrm{Hz}$.

\subsection{Resolved events}

An advantage of using the characteristic strain to describe the amplitude of GW sources given the sensitivity of the detector is that the S/N averaged over all possible orientations of the source and interferometer can be written as
\begin{equation}\label{eq:S/N}
	\mathrm{S/N} = \sqrt{ \int _f^{f+\Delta f} \left[ \frac{h_c(f_{\mathrm{obs}}^\prime)}{h_n(f_{\mathrm{obs}}^\prime)}	\right]^2 \frac{df_{\mathrm{obs}}^\prime}{f_{\mathrm{obs}}^\prime}},
\end{equation}
which allows one to assess by eye the detectability of a given source if $h_c$ is plotted against the observed frequency \citep{Moore:2015}.

The resolution frequency bin, $\Delta f$, is set by the minimum frequency resolvable by the instrumentation. It is the inverse of the mission lifetime $\Delta f \sim 1/T_\mathrm{obs}$, where $T_\mathrm{obs}$ is the length of observation \citep{Gair:2013}. For small $\Delta f$, we can assume a constant ratio $k = h_c(f_{\mathrm{obs}}^\prime)/h_n(f_{\mathrm{obs}}^\prime)$. Then, by changing the integration limits as $f+\Delta f = f(1+\alpha)$, where $\alpha = \Delta f / f$, we can rewrite \cref{eq:S/N} as
\begin{equation}\label{eq:S/N_sim}
\begin{aligned}
	(\mathrm{S/N})^2 &=  k^2 \int _f^{f(1+\alpha)} \frac{df_{\mathrm{obs}}^\prime}{f_{\mathrm{obs}}^\prime} \\
		  &=  k^2 \ln (1+\alpha).
\end{aligned}
\end{equation}

The eLISA mission has an expected duration of 3 years. Therefore $\Delta f \sim  1/T_\mathrm{obs} = 1/(3 \mathrm{yrs}) \approx 10 ^ {-8} \, \mathrm{Hz}$. If we impose $S/N \gtrsim 5$ for all the frequency bandwidth $(3\times10^{-5}\,\mathrm{Hz}$ to $1\,\mathrm{Hz})$ in \cref{eq:S/N_sim}, this results in $k \gtrsim 1.76$. In the examples shown in \cref{fig:spectrum} it can be seen that once any given GW signal crosses the detector sensitivity curve, the ratio of the signal to the sensitivity curve, $k$, rapidly increases by a few orders of magnitude. Therefore we can safely assume that all GW signals above the sensitivity curve (i.e. $h_c(f_{\mathrm{obs}}) \geq h_n(f_{\mathrm{obs}})$) can be detected by eLISA.

\subsection{Event rate}

We calculate the number of detected sources (i.e. ${h_c(f_{\mathrm{obs}}) \geq h_n(f_{\mathrm{obs}})}$) per redshift interval $z + \Delta z$ and co-moving volume $V_c$, and denote this quantity as ${\bar{N}(z,k\geq 1)}/{\Delta z V_c} \approx {d^2\bar{n}(z,k\geq 1)}/{dzdV_c}$. Integrating over all redshifts, the estimated event rate of detected GW sources per observed time is given by
\begin{equation}\label{eq:event_rate}
	\frac{d\bar{N}}{dt_{\mathrm{obs}}} = \int^\infty_0 \frac{d^2\bar{n}(z,k\geq 1)}{dzdV_c}\frac{dz}{dt}\frac{dV_c}{dz}\frac{dz}{(1+z)}.
\end{equation}

The total number of observed events in a given observation time is simply
\begin{equation}\label{eq:total_event_rate}
N_\mathrm{total} = \int_0^{T_\mathrm{obs}} \frac{d\bar{N}}{dt_{\mathrm{obs}}} dt_{\mathrm{obs}},
\end{equation}
where $T_{\mathrm{obs}} = 3 \, \mathrm{yrs}$ is the length of the mission. We now seek to estimate this quantity using the merger rates of SMBHs in the \textsc{eagle} cosmological hydrodynamical simulations.  

\begin{table*}
\begin{tabular}{cccccccc}\hline
\multicolumn{1}{l}{Simulation} & L & N & ${m}_{\mathrm{gas}}$ & ${m}_{\mathrm{DM}}$ & $\epsilon_{\mathrm{com}}$ &    $\epsilon_{\mathrm{prop}}$ & $m_\mathrm{seed}$ \\
 & $[\mathrm{cMpc}]$ &  & $[\mathrm{M}_{\sun}]$  & $[\mathrm{M}_{\sun}]$ & $[\mathrm{ckpc}]$ & $[\mathrm{ckpc}]$ & $[\mathrm{M}_{\sun}]$\\ \hline
\multicolumn{1}{l}{Ref-L100N1504} & $100$ & \multicolumn{1}{r}{$2 \times 1504^3$} & $1.81 \times 10^6$ & $9.70 \times 10^6$ & $2.66$   & $0.70$  & $1.475 \times 10^5$ \\
\multicolumn{1}{l}{SS-L050N0752} & $\mathbf{50}$ & \multicolumn{1}{r}{$2 \times 752^3$}  & $1.81 \times 10^6$ & $9.70 \times 10^6$ & $2.66$   & $0.70$ & $\mathbf{1.475 \times 10^4}$\\ 
\multicolumn{1}{l}{Ref-L050N0752} & $\mathbf{50}$ & \multicolumn{1}{r}{$2 \times 752^3$} & $1.81 \times 10^6$ & $9.70 \times 10^6$ & $2.66$ & $0.70$ & $1.475 \times 10^5$\\
\multicolumn{1}{l}{Ref-L025N0376} & $\mathbf{25}$ & \multicolumn{1}{r}{$2 \times 376^3$}  & $1.81 \times 10^6$ & $9.70 \times 10^6$ & $2.66$ & $0.70$  & $1.475 \times 10^5$\\
\multicolumn{1}{l}{Recal-L025N0752} & $\mathbf{25}$ & \multicolumn{1}{r}{$2 \times 752^3$}  & $\mathbf{2.26 \times 10^5}$ & $\mathbf{1.21 \times 10^6}$ & $\mathbf{1.33}$   & $\mathbf{0.35}$ &$1.475 \times 10^5$
\end{tabular}
\caption{Box-size, number of particles, initial baryonic and dark matter particle mass, co-moving and Plummer-equivalent gravitational softening, and BH seed mass for the \textsc{eagle} simulations used in this paper. Values in bold show differences with respect to the Ref-L100N1504 simulation.}\label{tab:a}
\end{table*}

\section{The EAGLE Simulations}\label{sec:Sim}

The \textsc{eagle} simulation suite\footnote{\url{http://www.eaglesim.org}} \citep{Schaye:2015,Crain:2015} consists of a large number of cosmological hydrodynamical simulations that include different resolutions, simulated volumes and physical models. These simulations use advanced smoothed particle hydrodynamics (SPH) and state-of-the-art subgrid models to capture the unresolved physics. Radiative cooling \citep{Wiersma:2009Cooling}, star formation \citep{SchayeDallaVecchia:2008,Schaye:2004}, metal enrichment \citep{Wiersma:2009Enrichment}, energy input from stellar feedback \citep{DallaVecchiaSchaye:2012}, BH growth \citep{Rosas-Guevara:2015,Schaye:2015}, and feedback from active galactic nuclei \citep{Schaye:2015} are included. The simulation suite was run with a modified version of the \textsc{gadget-}${\scriptstyle 3}$  SPH code (last described by \citealt{GADGET}) and includes a full treatment of gravity and hydrodynamics. The modifications to the SPH method, collectively referred to as \textsc{anarchy} (Dalla Vecchia et al. in preparation), make use of the pressure-entropy formulation of SPH derived by \cite{Hopkins:2013}, the artificial viscosity switch from \cite{Cullen:2010}, an artificial conduction switch similar to that of \cite{Price:2008}, the $\mathcal{C}_2$ kernel of \cite{Wendland:1995}, and the time-step limiters of \cite{Durier:2012}. The effects of this state-of-the-art formulation of SPH on the galaxy properties are explored in detail by \cite{Schaller:2015SPH}. The calibration strategy is described in detail by \cite{Crain:2015} who also presented additional simulations to demonstrate the effect of parameter variations. 

The halo and galaxy catalogues for more than $10^5$ simulated galaxies of the main \textsc{eagle} simulations with integrated quantities describing the galaxies, such as stellar mass, star formation rates, metallicities and luminosities, are available in the \textsc{eagle} database\footnote{\url{http://www.eaglesim.org/database.php}} \citep{McAlpine:2015-DB}. A complete description of the code and physical parameters used in the \textsc{eagle} simulations can be found in \citet{Schaye:2015}, here we present a brief overview of the BH seeding and growth mechanisms. Cosmological parameters for a standard $\Lambda$CDM universe were adopted by these simulations. The values of the key cosmological parameters implemented are $ \Omega_\mathrm{m} = 0.307$, $ \Omega_\Lambda = 0.693$, $\Omega_\mathrm{b} = 0.04825$, $h = 0.6777$ and $\sigma_8 = 0.8288$, as inferred by the \citet{Planck}. 

The label for each simulation denotes the comoving cubic box length and the cube root of the number of particles in the simulation. For example, Ref-L100N1504 corresponds to a simulation volume of $(100 \, \mathrm{cMpc})^3$ (where $\mathrm{cMpc}$ denotes co-moving megaparsecs) using $1504^3$ particles of dark matter and an equal number of baryonic particles. A prefix distinguishes the subgrid variations. For example, the prefix \enquote*{\textit{Ref-}} refers to a simulation using the reference model. 

We compare the predicted GW signals from two \textsc{eagle} models, our reference simulation Ref-L100N1504, and a modified version of the Ref-L050N0752 model which uses the same calibrated subgrid parameters as the reference model, but smaller BH seeds, as described in \cref{sec:BHSeeds}. We have labeled this model SS-L050N0752. Additionally, in order to test for convergence with simulated volume size and resolution, the Ref-L050N0752, Ref-L025N0376 and Recal-L025N0752 models are compared in \cref{sec:App1}. As discussed by \citet{Schaye:2015}, the \enquote*{\textit{Recal-}} higher-resolution simulation uses values of the subgrid parameters that were recalibrated following the same procedure used for the reference simulation, enabling the user to test the ``weak convergence'' properties of the simulations. In \cref{tab:a} we summarise the simulation models used in this paper, including the comoving cubic box length, initial baryonic and non-baryonic particle masses, Plummer-equivalent gravitational softening lengths and BH seed mass. Together these parameters determine the dynamic range and resolution that can be achieved by the simulations.

\subsection{Black hole seeding}\label{sec:BHSeeds}
To explain the population of luminous quasars in the high-redshift Universe $(z \geq 6 )$ SMBHs must have formed early and grown rapidly (e.g. \citealt{VolonteriBellovary:2012}). Different formation and evolution mechanisms for BH seeds have been proposed  to explain the rapid growth that enables  these seeds to grow to masses of $10^9 \, \mathrm{M}_{\sun}$ in less than one billion years. These SMBHs may have originated from the remnants of the very first generation of stars, runaway collisions of stars and/or stellar mass BHs, direct collapse of supermassive stars, or from an even more exotic process (refer to \citealt{Volonteri:2010} for a review on formation models for SMBHs). We now briefly review the most promising models for forming SMBH seeds in the early Universe:

\begin{enumerate}
	\item \textit{Remnants of the first generation of stars}
	
BH seeds may have formed from remnants of Population III stars \citep[e.g.][]{Madau.and.Rees:2001,VHM:2003}. If stars more massive than $\sim 250 \, \mathrm{M}_{\sun}$ formed from primordial gas, they are predicted to directly collapse into BH seeds with masses of $\sim 100 \, \mathrm{M}_{\sun}$. However, it is still unclear if most of the first stars were born with such large masses \citep[e.g.][]{Clark:2011,Greif:2011}. Additionally, in order to grow to masses in excess of $10^9 \, \mathrm{M}_{\sun}$ as early as redshift $z \sim 6$ seeds would require to grow close to the Eddington rate for the majority of their lifetime. The shallow potential wells in which Population III stars form makes this scenario rather unattractive \citep[e.g.][]{Johnson:2008,Alvarez:2009,Volonteri:2015edd}. Growth through super-Eddington accretion phases may solve this conflict \citep[e.g.][]{Volonteri:2015Eddington,Lupi:2016,Inayoshi:2016}. However, further theoretical work on this mechanism and its sustainability is required. 
	
	\item \textit{Collapsing nuclear star clusters}
	
In this model, stellar-dynamical instabilities in proto-galactic discs may lead to infall without fragmentation of low metallicity gas, increasing the central galactic density \citep[e.g.][]{Seth:2008,Devecchi:2009,DevecchiVolo:2012,Lupi:2014,Katz:2015}. Within the nuclear region a dense stellar cluster forms. As the central cluster undergoes core collapse, runaway collisions of stars may lead to the formation of a single SMBH seed with a mass up to $\sim 10^3 \, \mathrm{M}_{\sun}$.
	
	\item \textit{Direct collapse of supermassive stars}
	
It has been proposed that in high-redshift haloes radiation emitted by nearby star-forming galaxies could cause the photo-dissociation of $\mathrm{H}_2$. This prevents the temperature of primordial gas from reaching very low values and thus elevates the Jeans mass, allowing the formation of a large central mass, possibly evolving into a supermassive star \citep[e.g.][]{Omukai:2001,Wise:2008,Regan:2009,Agarwal:2014,Sugimura:2014,Regan:2015}.	
Another mechanism to form a supermassive star is by rapid funnelling of low metallicity gas in low angular momentum haloes with global or local dynamical instabilities \citep[e.g.][]{Loeb:1994,KBD:2004,Begelman:2008}. Once a supermassive star forms its core may collapse to form a small BH within the radiation-pressure-supported object. In this scenario the central BH can accrete the entire envelope and form a SMBH seed of mass $\sim 10^3 \, \mathrm{M}_{\sun}$ up to $\sim 10^5 \, \mathrm{M}_{\sun}$.
\end{enumerate}
	
Constraining the formation mechanisms of BH seeds represents a major observational challenge. As we will show, the detection of GW signals from SMBH coalescences represents a promising way to discriminate among different theoretical formation models by determining the mass function of seed BHs.

Since the SMBH seed formation processes are not resolved by cosmological simulations, it is assumed that every halo above a certain threshold mass hosts a central BH seed. For a comprehensive description of the BH seeding mechanisms in these simulations see \citealt{Springel:2005}, \citealt{DiMatteo:2008}, \citealt{Booth:2009}, and \citealt{Schaye:2015}. In the Ref-L100N1504 model high-mass BH seeds\footnote{$m_{\mathrm{seed}} = 1.475 \times 10^5 \mathrm{M}_{\sun} = 1 \times 10^5 \mathrm{M}_{\sun} h^{-1}$, where $h = 0.6777$.} $(m_{\mathrm{seed}} = 1.475 \times 10^5 \mathrm{M}_{\sun})$ are placed at the centre of every halo with total mass greater than $m_{\mathrm{halo,th}} = 1.475 \times 10^{10} \mathrm{M}_{\sun}$ that does not already contain a BH. For the SS-L050N0752 model the BH seed mass is $m_{\mathrm{seed}} = 1.475 \times 10^4 \mathrm{M}_{\sun}$. We choose to place BH seeds in haloes of mass $m_{\mathrm{halo,th}} = 1.475 \times 10^{10} \, \mathrm{M}_{\sun}$ (which corresponds to $m_{\mathrm{halo,th}} \sim 1.5 \times 10^3 \, m_{\mathrm{DM}}$ for the reference models and $m_{\mathrm{halo,th}} \sim 1.2 \times 10^4 \, m_{\mathrm{DM}}$ for the high-resolution \textit{`Recal'} model) to ensure that the structure of haloes containing BHs is always well resolved. Halos are selected for seeding by regularly running the ``Friends-of-Friends'' (FoF) halo finder on the dark matter distribution, with a linking length equal to $0.2$ times the  mean inter-particle spacing. 

\subsection{Black hole dynamics and delays}\label{sec:Delays}

Our aim is to calculate the GW signals from the merger rates of SMBHs across cosmic time, which depend crucially on how many galaxies host BHs and on the galaxy merger history. Therefore, full cosmological models including BH physics are necessary to study the merger rates of SMBHs. Nonetheless, given the limited spatial resolution in large scale cosmological simulations, information on the small-scale dynamical evolution of SBMH binaries is lost. To overcome this limit and obtain realistic SMBH dynamics and merger rates we employ advection schemes that correct the motion of BH particles and apply a time delay corrections to the BH merger timescales to account for the unresolved small-scale dynamical evolution of the binaries.

In the simulations, when a halo grows above the threshold mass $m_{\mathrm{halo,th}}$, its highest-density  gas particle is converted into a collisionless BH particle with subgrid mass $m_{\mathrm{BH}} = m_{\mathrm{seed}}$. Since the BH seed mass is usually significantly lower than the baryonic particle mass $(m_{\mathrm{seed}} \ll m_{\mathrm{gas}})$, the use of a subgrid mass is necessary for BH-specific processes such as accretion \citep{Springel:2005}. On the other hand, gravitational interactions are computed using the BH particle mass. 

Since the simulations cannot model the dynamical friction acting on BHs with masses $\lesssim m_{\mathrm{gas}}$, BHs with mass $<100$ times the initial gas particle mass $m_\mathrm{gas}$ are re-positioned to the local potential minimum. To prevent BHs in gas poor haloes from jumping to nearby satellites, we limit this process to particles whose velocity relative to the BH is smaller than $0.25c_\mathrm{s}$, where $c_\mathrm{s}$ is the local sound speed, and whose distance is smaller than three gravitational softening lengths. Tracking the evolution of BH orbits in individual galaxy mergers during code development showed this eliminated spurious repositioning events in fly-by encounters. However, within $\sim$kpc separations, repositioning of BHs to account for unrealistic dynamics in cosmological scale simulations may cause spurious SMBH mergers and/or SMBHs to merge sooner than what is predicted by their orbital decay time-scale \citep{Tremmel:2015}. Hence, a SMBH merger time delay is needed to correct for this effect.

Two BHs merge if they are separated by a distance that is smaller than both the SPH smoothing kernel of the BH, $h_{\mathrm{BH}}$, and three gravitational softening lengths (this criteria gives a median separation of $\sim 1 \mathrm{pkpc}$ at all redshifts and halo masses). $h_{\mathrm{BH}}$ is chosen such that within a distance $h_{\mathrm{BH}}$ from the BH there are $N_\mathrm{ngb} = 58$ weighted SPH neighbours. Furthermore, in order to prevent BHs from merging during fly-by encounters we impose a limit on the allowed relative velocity of the BHs, required to be smaller than the circular velocity at the distance $h_{\mathrm{BH}}$ ($v_{\mathrm{rel}} < \sqrt{G{M}_{1}h^{-1}_{\mathrm{BH}}}$, where $G$ is the gravitational constant and ${M}_{1}$ is the mass of the most massive BH in the pair). Triple BH mergers can happen in a single time-step in the simulations. However, due to their extreme rarity, we do not consider these events in our analysis.

As we briefly discussed in \cref{sec:GW}, after two galaxies merge a variety of effects can affect the dynamical evolution of the SMBH binary and finally lead (or not) to a merger within a Hubble time. Dynamical friction, three-body interactions with stars, interactions with gas, including planet-like migration and/or orbital decay of the binary due to the clumpy gas and the heating of the cold layer of the disc by BH feedback can either prevent or promote the SBMH merger \citep{Colpi:2011, Mayer:2013, Colpi:2014, Tamburello:2016}. A binary could stall at $\sim \mathrm{pc}$ separations, which is known as the ``final-parsec'' problem \citep{Begelman:1980}. However, a later galaxy merger may trigger the SMBH merger \citep{Hoffman:2007}. Given the uncertainties in these mechanisms, and the potential variability from galaxy to galaxy, we adopt simplified prescriptions to estimate the SMBH merger time delays based on the gas content in the nuclear region of the resulting galaxy after the merger, similar to the method adopted by \cite{Delays}.
\begin{itemize}
	\item \textit{Gas rich galaxies}: For galaxies with gas mass within a 3D aperture with radius 3 pkpc greater than or equal to the total BH mass $(M_{\mathrm{gas@3pkpc}} \geq M_1 + M_2)$, a $0.1 \, \mathrm{Gyr}$ delay was added to the BH merger time recorded in the simulation. \\
	\item \textit{Gas poor galaxies}: For galaxies with gas mass within a 3D aperture with radius 3 pkpc less than the total BH mass $(M_{\mathrm{gas@3pkpc}} < M_1 + M_2)$, a $5 \, \mathrm{Gyr}$ delay was added to the BH merger time recorded in the simulation.
\end{itemize}

As we will discuss in \cref{sec:results}, we find that adding a significant delay in gas-poor galaxies gives a very similar result of the expected rate of GW signals than our model assuming no delays.

 \begin{figure}\centering
 \includegraphics[width=0.45\textwidth]{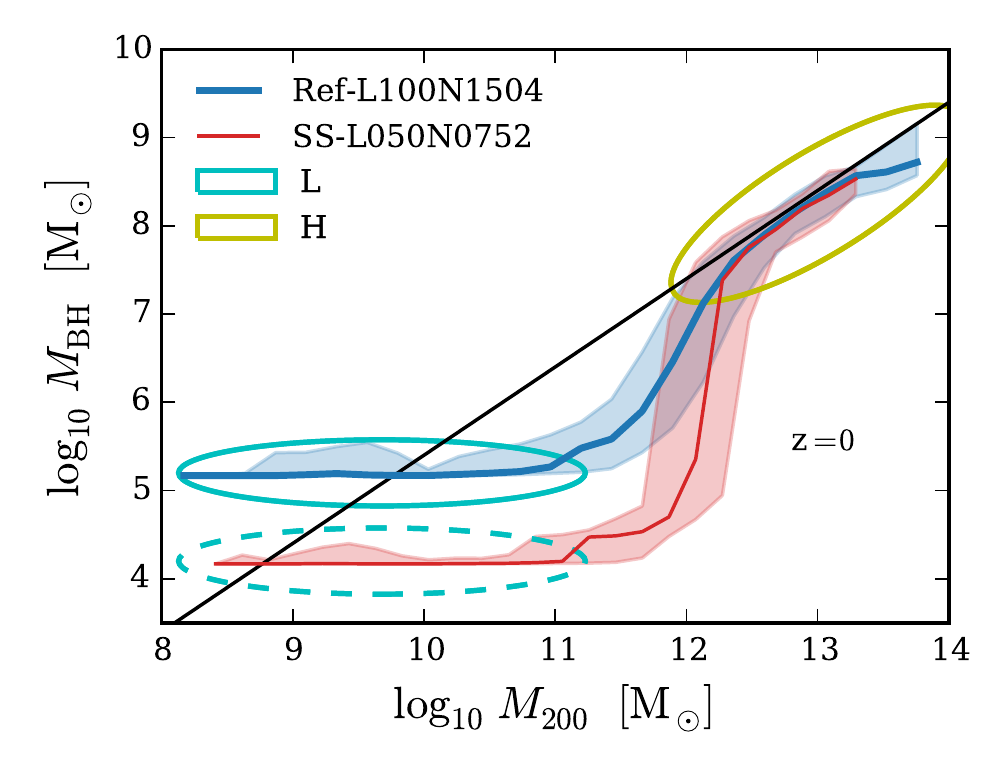}
 \vspace{-1.5em}
 \caption{Halo mass-central BH mass relation for two \textsc{eagle} simulations at redshift $z=0$. Lines represent the median of the distribution for each simulation. Only bins containing $5$ objects or more have been plotted. The shaded region encloses the $10^\mathrm{th}$ to $90^\mathrm{th}$ percentiles for each model. The black line is shown as reference for a relationship $\mathrm{M}_\mathrm{BH} \propto \mathrm{M}_{200}$. Regardless of the initial BH seed mass the halo mass-BH mass relation exhibits a steep slope in haloes with mass ${\sim}10^{12}\mathrm{M}_{\sun}$. At this halo mass, the hot gas in the corona causes the star formation driven outflows to stall and conditions become optimal for BH accretion, and BHs grow rapidly (Bower et al. in preparation). For BHs hosted by haloes more massive than  ${\sim}10^{12}\mathrm{M}_{\sun}$ the growth is self-regulated by AGN feedback. Two prominent populations of SMBHs are highlighted: BHs not much more massive than the seed mass (\textbf{L} for \enquote*{low mass}) and very massive BHs with masses $>10^7$ $\mathrm{M}_{\sun}$ (\textbf{H} for \enquote*{high mass}).}
 \label{fig:BH_Halo}
\end{figure}

 \begin{figure*}\centering
 \includegraphics[width=0.99\textwidth]{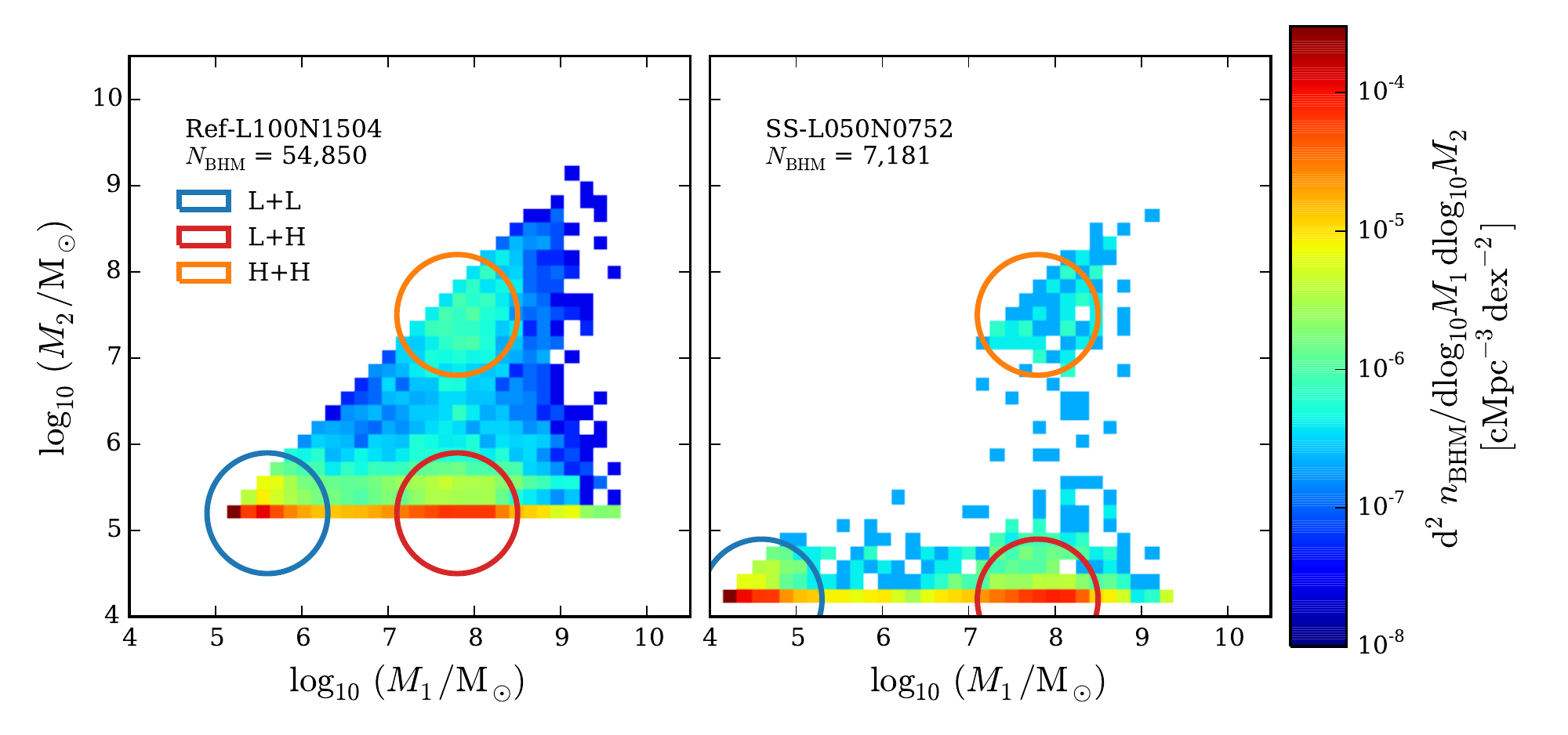}
 \vspace{-.5em}
 \caption{2D histogram of all BH mergers for all redshifts in the \textsc{eagle} simulations. Ref-L100N1504 (left panel) and SS-L050N0752 (right panel). ${M}_1$ is the more massive member of the SMBH binary (${M}_1 \geq {M}_2$). The total number of coalescence events in each simulation model, $N_{\mathrm{BHM}}$, is shown in the top left corner of each panel. Colour coding represents the number density of SMBH mergers per binary mass bin. As a result of the transition in the Halo mass-central BH mass relation shown in \cref{fig:BH_Halo}, there are three prominent populations of SMBH binaries in the simulations, which are highlighted in the figure: SMBH binaries that involve two BHs not much more massive than the seed mass (\textbf{L}$+$\textbf{L}); high mass ratio binaries where ${M}_1$ is massive ($>10^7 \mathrm{M}_{\sun}$) and ${M}_2$ is not much more massive than the seed mass (\textbf{L}$+$\textbf{H});  the case where both BHs are massive, with masses above $10^7$ $\mathrm{M}_{\sun}$ (\textbf{H}$+$\textbf{H}). Since the populations of high-mass SMBHs in both simulation models reach similar masses (\cref{fig:BH_Halo}), the population of (\textbf{H}$+$\textbf{H}) binaries occupies the same region in both models. On the other hand, both the (\textbf{L}$+$\textbf{L}) and (\textbf{L}$+$\textbf{H}) populations are shifted in $M_1$ and $M_2$ in the SS-L050N0752 model compared to Ref-L100N1504.}
 \label{fig:2Dmasshist}
\end{figure*}

\subsection{Black hole growth}\label{sec:BHGrowth}

Once seeded, BHs are free to grow via gas accretion at a rate that depends only on the local hydrodynamical properties, namely:  the mass of the BH, the local density and temperature of the surrounding gas, the velocity of the BH relative to the ambient gas, and the angular momentum of the gas with respect to the BH. Accretion onto BHs follows a modified version of the Bondi-Hoyle accretion rate which takes into account the circularisation and subsequent viscous transport of infalling material, limited by the Eddington rate (as described by \citealt{Rosas-Guevara:2015}). Additionally, BHs can grow by merging with other BHs as described in the previous section.

It is important to highlight that the sub-grid physics in the \textsc{eagle} simulations were calibrated to reproduce the observed galaxy stellar mass function at redshift $z=0.1$, the amplitude of the galaxy stellar mass-central BH mass relation and galaxy sizes \citep{Crain:2015}. Although not part of the calibration procedure, \citealt{Rosas-Guevara:2016} show that the simulations also reproduce the observed BH mass function at $z = 0$ and show good agreement with the observed AGN luminosity functions in the hard and soft X-ray bands. Additionally, \cite{Trayford:2016:Green} shows the important role of BH growth in quenching star formation and establishing the high-mass red sequence of galaxies in \textsc{eagle}.

\Cref{fig:BH_Halo} shows the halo mass-central BH mass relation at redshift $z = 0$ for the \textsc{eagle} simulations discussed here. The halo mass, ${M}_{200}$, is defined as the total mass within the radius within which the mean density is $200$ times the critical density of the Universe. 

Regardless of the initial seed mass, BHs that reside in low-mass haloes barely grow because star formation driven outflows are efficient and able to prevent large reservoirs of cold low angular momentum gas accumulating around the BH. Then, the accretion behaviour changes dramatically in haloes with mass $\sim10^{12}\mathrm{M}_{\sun}$. At this halo mass, the hot gas in the corona causes the star formation driven outflows to stall and conditions become optimal for BH accretion, and BHs grow rapidly. The growth of BHs residing in haloes more massive than  $\sim10^{12}\mathrm{M}_{\sun}$ is self-regulated by AGN feedback and BHs reach similar masses regardless of their initial seed mass. The physical origin of this transition is further discussed by Bower et al. (in preparation). As a result of this transition there are two prominent populations of SMBHs in the simulations. These are highlighted in the figure: BHs not much more massive than the seed mass (\textbf{L} for \enquote*{low mass}) and very massive BHs with masses $>10^7$ $\mathrm{M}_{\sun}$ (\textbf{H} for \enquote*{high mass}).

\begin{figure}
\hspace*{-4mm}
 \includegraphics[width=0.5\textwidth]{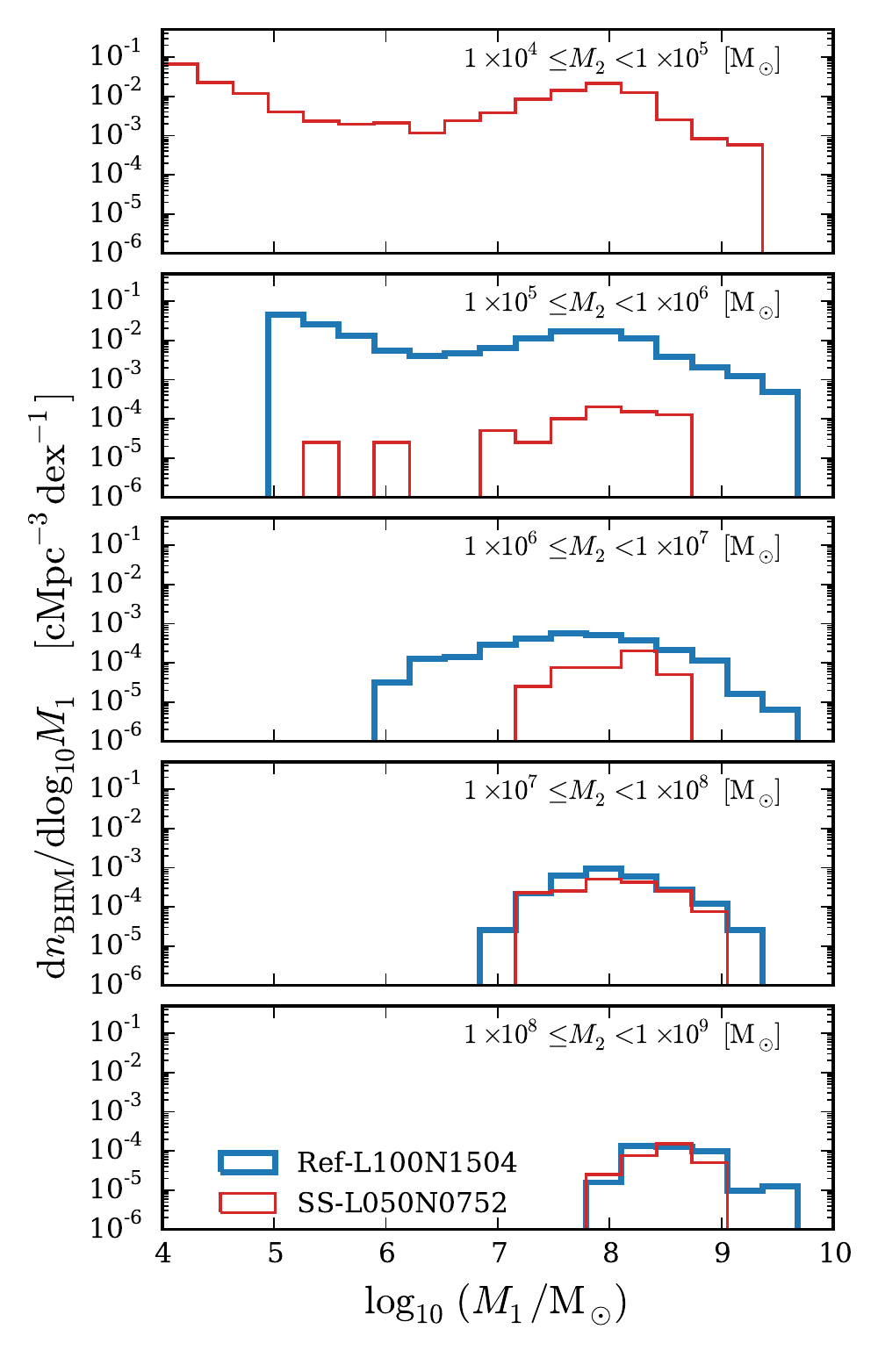}
 \vspace{-1.5em}
 \caption{Co-moving number density of the more massive member of the BH binaries with mass ${M}_1$, for five non-contiguous ranges in the mass of the least massive member, ${M}_2$, as indicated in the legend (top to bottom). The population of SMBH binaries that involve a BH not much more massive than the seed mass ($<10^6 \mathrm{M}_{\sun}$) are then shown in the top panel. Binaries where both BHs are massive ($>10^7 \mathrm{M}_{\sun}$) are shown in the two bottom panels. The population (\textbf{H}$+$\textbf{H}), shown in the two bottom panels, is at least two orders of magnitude smaller than the (\textbf{L}$+$\textbf{L}) and (\textbf{L}$+$\textbf{H}) populations, shown in the top two panels.}
 \label{fig:masspanels}
\end{figure}

\subsection{Black hole coalescence}

For each SMBH merger that takes place in the simulations we store the mass of both SMBHs, $M_1$ and $M_2$, and the redshift $z$ at which  the merger event takes place. On \cref{fig:2Dmasshist} we show the 2D histogram of the mass of each BH member for all the mergers in the \textsc{eagle} simulation models considered here. The total number of BH mergers in each simulation model is indicated in the figure. For Ref-L100N1504 a total of $N_\mathrm{BHM}=54,850$ BH mergers take place across cosmic time. A factor of $\approx 8$  fewer mergers occur in the small seeds ($N_\mathrm{BHM}=7,045$) model almost entirely due to the factor of 8 smaller volume of the simulation. Three prominent populations of characteristic SMBH binaries build up. These are the result of the halo mass-central BH mass relation in \textsc{eagle}, shown in \cref{fig:BH_Halo}. The groups are: SMBH binaries that involve two BHs not much more massive than the seed mass (\textbf{L}$+$\textbf{L}); high mass ratio binaries where ${M}_1$ is massive ($>10^7 \mathrm{M}_{\sun}$) and ${M}_2$ is not much more massive than the seed mass (\textbf{L}$+$\textbf{H}); the case where both BHs are massive, with masses between $10^7$ and $10^8$ $\mathrm{M}_{\sun}$ (\textbf{H}$+$\textbf{H}). 
 
In \cref{fig:masspanels} we show the co-moving number density distribution as a function of the more massive member of the BH binaries, ${M}_1$, plotted for five non-contiguous ranges in the mass of the least massive member, ${M}_2$. The (\textbf{L}$+$\textbf{L}) and (\textbf{L}$+$\textbf{H}) populations of binaries correspond to the left and right peaks of the distribution in the top two panels. The population (\textbf{H}$+$\textbf{H}) is shown in the bottom two panels. Naturally, the larger simulation volume samples more massive structures, hence the observed BH coalescence distribution in each panel extends to higher values of $M_1$. Since we keep the same vertical axis range for all five panels we can compare the contribution of each mass bin to the total SMBH merger rate. The population of binaries where both BHs are massive (\textbf{H}$+$\textbf{H}), shown in the bottom panel, is at least two orders of magnitude smaller than that of high mass ratio binaries (\textbf{L}$+$\textbf{H}) and SMBH binaries that involve two BHs not much more massive than the seed mass (\textbf{L}$+$\textbf{L}), shown in the top two panels of the figure.

\begin{figure}\centering
 \includegraphics[width=0.45\textwidth]{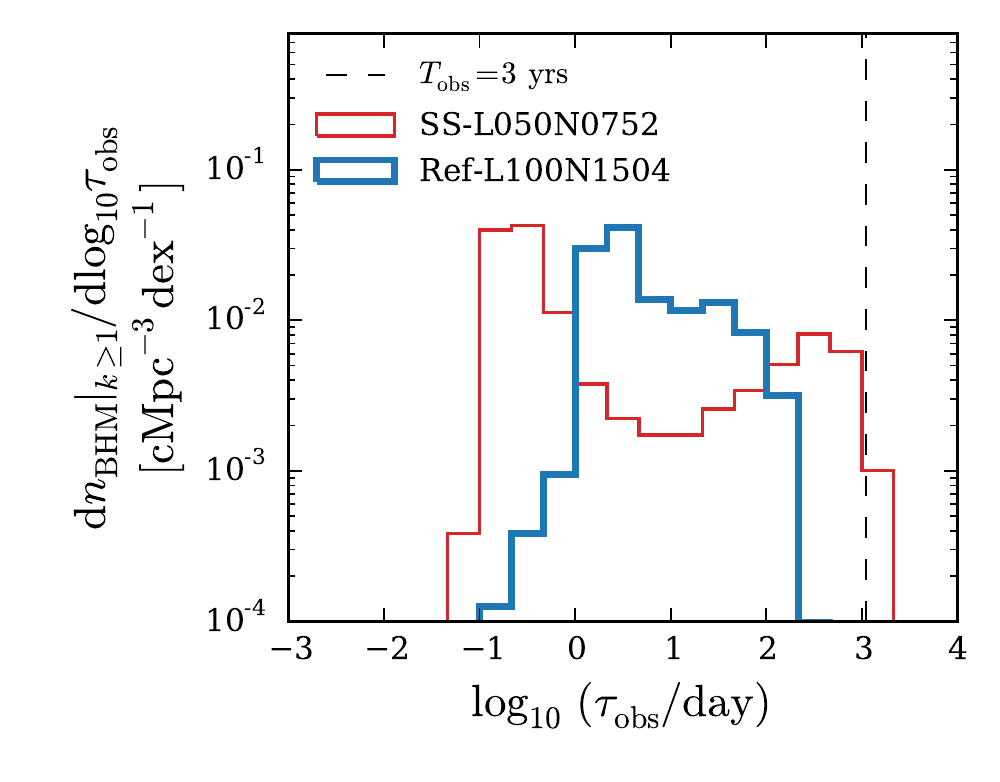}
 \vspace{-1.5em}
 \caption{Distribution of the observed duration of the events that would be resolvable by the eLISA detector (i.e. the ratio of the GW signal to the detector's sensitivity curve $k\geq1$).  For this study, we assume an eLISA mission lifetime of $T_\mathrm{obs} = 3 \, \mathrm{yrs}$ (dashed vertical line). We only consider events with $\tau_\mathrm{obs} \leq T_\mathrm{obs}$ in the rest of the paper.}
 \label{fig:tau}
\end{figure}

\begin{figure*}
 \includegraphics[width=0.99\textwidth]{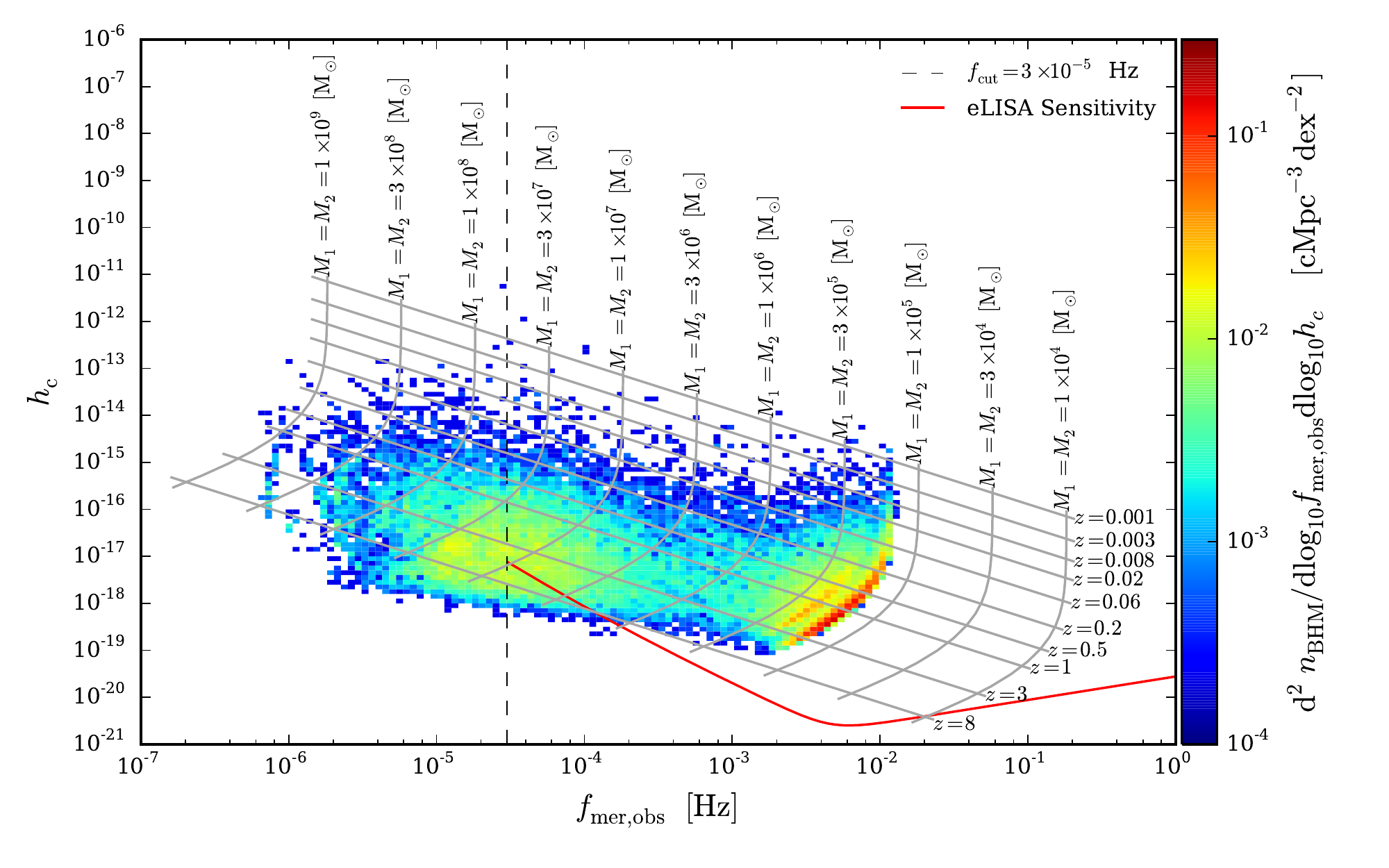}
 \vspace{-1.5em}
 \caption{Characteristic strain amplitude $h_c$ of the GW signals emitted by all SMBH coalescences in the \textsc{eagle} Ref-L100N1504 simulation as a function of the observed frequency at the transition between the inspiral phase and the merger phase of the SBMH coalescence process $f_\mathrm{mer,obs} = (0.018 \, c^3/GM_{\mathrm{total}})/(1+z)$. Colour coding represents the co-moving number density of events per characteristic strain-observed merger frequency bin. Grey contour lines indicate the characteristic strain and observed merger frequency for equal mass BH binaries (${M}_1={M}_2$) coalescing at different redshifts $z$. The sensitivity curve of eLISA calculated from the analytic approximation in \cref{eq:sensitivity} is shown in red. The black dashed line indicates the low-frequency cut-off of the sensitivity curve $f_\mathrm{cut} = 3 \times 10 ^{-5} \,\, \mathrm{Hz}$. GW signals above the sensitivity curve and to the right of the low-frequency cut-off can be resolved from the eLISA data stream.}
 \label{fig:freq-hc}
\end{figure*}

\begin{figure*}
 \includegraphics[width=0.99\textwidth]{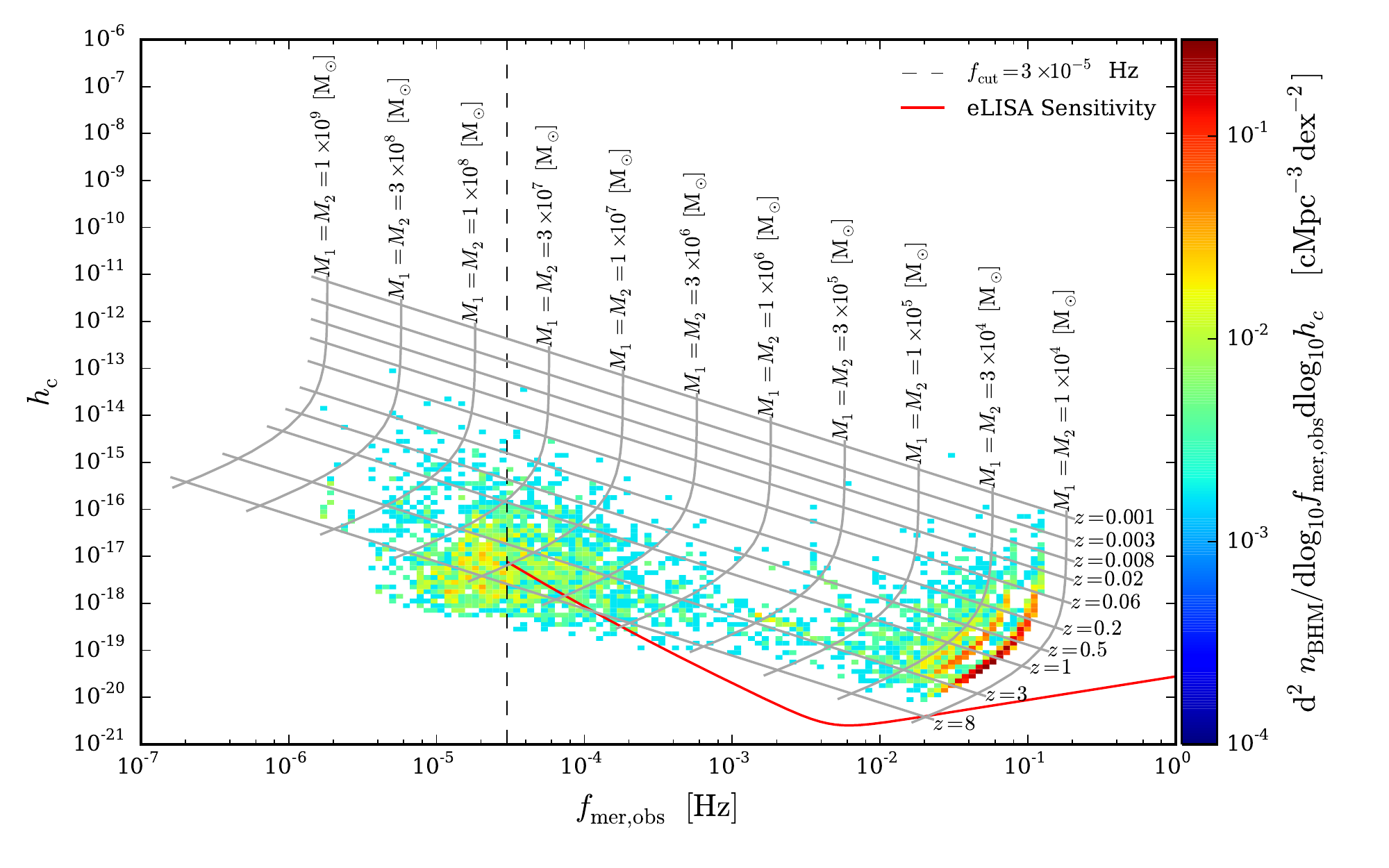}
 \vspace{-1.5em}
 \caption{As \cref{fig:freq-hc} but for the \textsc{eagle} SS-L050N0752 simulation.}
 \label{fig:freq-hc2}
\end{figure*}

\begin{figure}\centering
 \includegraphics[width=0.45\textwidth]{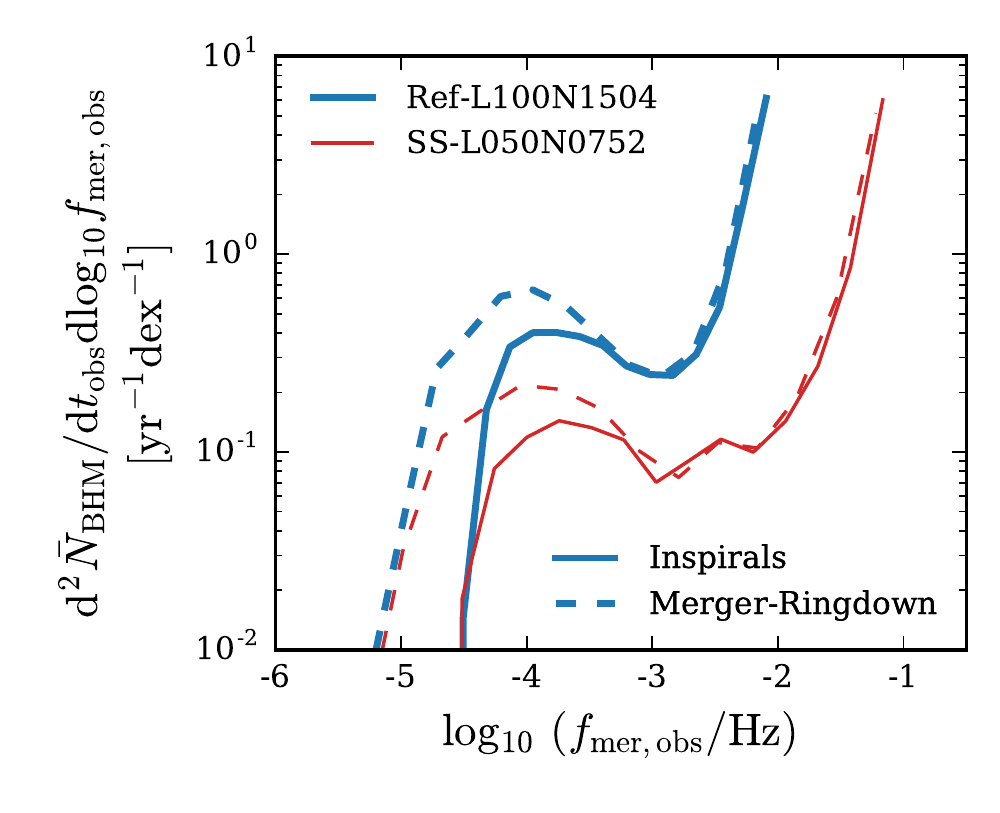}
 \vspace{-1.5em}
 \caption{Number of detected SMBH coalescences per observed year as a function of the frequency at the transition between the inspiral phase and the merger phase $f_\mathrm{mer,obs}$. A shift of a decade in frequency of the whole distribution is observed for the SS-L050N0752 compared to the Ref-L100N1504 model. The amplitude of the peak of the distribution is the same for both models. For lower frequencies (i.e. more massive mergers), the SS-L050N0752  model has fewer detected events (by about $0.3 \, \mathrm{dex}$) because these events have lower characteristic strain amplitude, as expected for lower mass BHs, and therefore fall outside the detection threshold of the eLISA sensitivity curve. }
 \label{fig:f_shift}
\end{figure}

\begin{figure*}
 \includegraphics[width=1\textwidth]{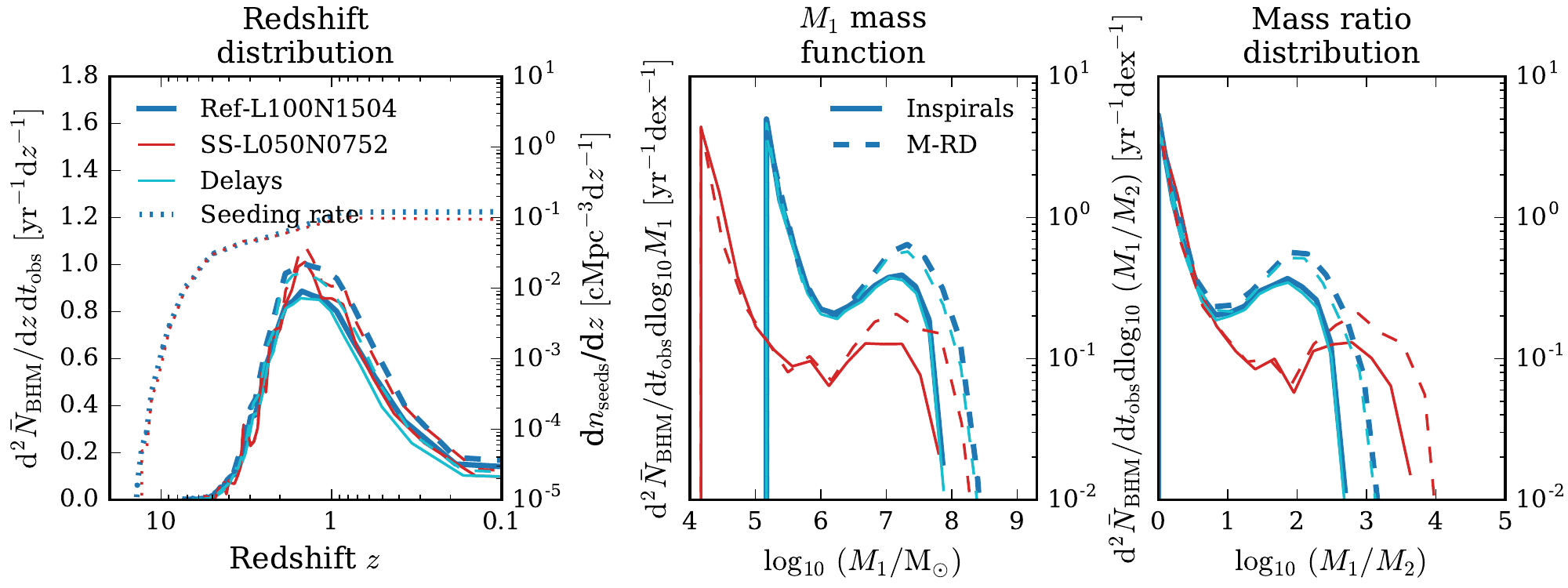}
 \vspace{-2em}
 \caption{\textit{LEFT PANEL:} Redshift distribution of the SMBH coalescences resolved by eLISA. The distribution peaks between redshift $z\sim2$ and $z\sim1$ for both the inspiral and the merger-ringdown phases for both \textsc{eagle} simulation. The differential redshift distribution of the SMBH seeding times are shown in doted lines (right-hand axis). \textit{MIDDLE PANEL:} Mass distribution of the more massive member of the binary, ${M}_1$, for the SMBH coalescences resolved by eLISA. For both models, the mass function peaks at ${M}_1 \sim m_{\mathrm{seed}}$ for both the inspiral and the merger-ringdown phases. \textit{RIGHT PANEL:} Distribution of the mass ratio ${M}_1/{M}_2$ of the SMBH coalescences resolved by eLISA. The distribution peaks for equal mass SMBH binaries for both \textsc{eagle} simulation models. The model with added delay to the SMBH merger timescales has no significant difference in either the predicted event rate, nor in the mass function of the detected binaries.}
 \label{fig:Results}
\end{figure*}

\section{Predicted gravitational wave event rate}\label{sec:results}

In order to compute the expected GW signals from SMBH mergers in the simulations, we adopt the following strategy. We first calculate the frequency at the transition from the inspiral phase to the merger phase, $f_{\mathrm{merger}} = 0.018 \, c^3/GM_{\mathrm{total}}$, for each merger event in \textsc{eagle}. We compute the minimum inspiral frequency $f_{\mathrm{min}} = 1\times 10^{-3} \, c^3/GM_{\mathrm{total}}$ and an arbitrary final frequency $f_{\mathrm{max}} = 2 \times 10^{-1} \, c^3/GM_{\mathrm{total}}$. Assuming that both SMBHs in the binary have no spin, we use the Inspiral-Merger-Ringdown waveform model PhenomD \citep{PhenomD2} to compute the characteristic strain amplitude of the GWs from each binary, which depends only on the merging redshift ($z$) and the mass of each SMBH ($M_1$ and $M_2$).
	We compute the sensitivity curve of eLISA using the analytical approximation given in \crefrange{eq:sensitivity}{eq:noise3}. For this analysis we adopt the target frequency cut of the detector, $f_{\mathrm{cut}} = 3 \times 10^{-5} \,\, \mathrm{Hz}$.
	From all the computed GW signals, we filter the events that would be resolvable by the detector using \cref{eq:S/N_sim} with a ratio of the signal to the sensitivity curve $k\geq1$ (i.e. ${\mathrm{S}/\mathrm{N} \gtrsim 5}$).
	For all resolvable events we compute the observed duration $\tau_\mathrm{obs}=\tau_\mathrm{inspiral,obs}+\tau_\mathrm{merger,obs}$ using \cref{eq:t_ins,eq:t_mer} for the detected frequencies. 
Finally, from the number of resolvable sources we estimate the event rate of GW sources and the total number of expected observable events during the lifetime of the eLISA mission using \cref{eq:event_rate,eq:total_event_rate}.   

For SMBH binaries the S/N of the GW signals is accumulated in the last month of the inspiral phase \citep{SesanaGair:2011}. Therefore, in this study we choose to only include sources that ``merge'' during the mission time, i.e., to construct the expected event rate we only consider events with $\tau_\mathrm{obs} \leq T_\mathrm{obs}$. In \cref{fig:tau} we show the distribution of the observed duration for all the resolvable events. The performance of the detector improves as a function of the duration of mission and gaps in the data stream would affect the number of resolved events \citep[e.g.][]{SesanaGair:2011}. For this study we assume a fiducial eLISA mission continuous lifetime of $T_\mathrm{obs} = 3 \mathrm{yrs}$. We show in \cref{fig:freq-hc,fig:freq-hc2} the characteristic strain amplitude as a function of the observed merger frequency for all the GW events produced by  SMBH coalescences in the \textsc{eagle} Ref-L100N1504 and SS-L050N0752 simulations respectively. To help visualise the mass range and redshift of BH coalescences that would be detected by eLISA, grey lines indicate the characteristic strain and observed merger frequency emitted by equal mass BH binaries (${M}_1={M}_2$) coalescing at different redshifts $z$. The following characteristic features can be seen in the figures: 

\begin{enumerate}
\item For both models, the most densely populated region of events (${1 \times 10 ^{-3} \lesssim f_\mathrm{mer,obs}/[\mathrm{Hz}] \lesssim 1 \times 10 ^{-2}}$ and ${1 \times 10 ^{-19} \lesssim h_c \lesssim 1 \times 10 ^{-17}}$ for Ref-L100N1504 and ${1 \times 10 ^{-2} \lesssim f_\mathrm{mer,obs}/[\mathrm{Hz}] \lesssim 1 \times 10 ^{-1}}$ and ${1 \times 10 ^{-20} \lesssim h_c \lesssim 1 \times 10 ^{-18}}$ for SS-L050N0752) corresponds to SMBH binaries where both BHs have masses not much greater that the seed mass (\textbf{L}$+$\textbf{L}). As illustrated in \cref{fig:spectrum}, we expect higher GW frequencies and smaller strain amplitudes from less massive BH mergers. Therefore, for the small seeds model the (\textbf{L}$+$\textbf{L}) population shifts to higher frequencies and lower amplitudes compared to the reference model. For both Ref-L100N1504 and SS-L050N0752 \textsc{eagle} models the (\textbf{L}$+$\textbf{L}) population of events occupies a region above the detection threshold of eLISA and hence will provide a high contribution to the data stream (as seen in \cref{fig:Results}).
\item The second most populated region of events (${1 \times 10 ^{-5} \lesssim f_\mathrm{mer,obs}/[\mathrm{Hz}] \lesssim 1 \times 10 ^{-4}}$ and ${1 \times 10 ^{-18} \lesssim h_c \lesssim 1 \times 10 ^{-16}}$ for both simulations) corresponds to binaries from the (\textbf{L}$+$\textbf{H}) population. For both the Ref-L100N1504 and SS-L050N0752 \textsc{eagle} models, there are significantly less events in  the (\textbf{L}$+$\textbf{H}) population compared to  the (\textbf{H}$+$\textbf{H}), with some falling outside the detection threshold of eLISA. Hence will not contribute significantly to the data stream.
\item Only few events from the (\textbf{H}$+$\textbf{H}) population occupy a region above the detection threshold of eLISA. However, there are significantly fewer events in the in this population of binaries compared to the (\textbf{L}$+$\textbf{L}) and (\textbf{L}$+$\textbf{H}) populations (at least two orders of magnitude fewer events, as seen in \cref{fig:masspanels}). Therefore, the binaries from the (\textbf{H}$+$\textbf{H}) population do not show up as a dense region in the plot.  
\end{enumerate}

The characteristic shift in amplitude and frequency of the detected GW signals that results from the different SMBH seed masses suggests that eLISA will be a powerful tool to discriminate between different SMBH seeding mechanisms. In \cref{fig:f_shift} we show the number of detected SMBH coalescences observed per year at redshift $z = 0$ as a function of the frequency at the transition between the inspiral phase and the merger phase, $f_\mathrm{mer,obs}$. The figure shows a shift of a decade in frequency for the whole distribution between the simulation models. The amplitude of the peak of the distribution is, however, the same for both models. For lower frequencies (i.e. more massive mergers) the SS-L050N0752  model has fewer detected events (${\sim}0.3 \, \mathrm{dex}$) because these events have lower characteristic strain amplitude and therefore some fall below the detection threshold of the eLISA sensitivity curve. 

\begin{table*}
\centering
\begin{tabular}{lccccl}
\hline
\multicolumn{1}{c}{Simulation} & Inspiral Phase      & $\sigma_\mathrm{I}$    & Merger-Ringdown Phase & $\sigma_\mathrm{M-RD}$    &  \\
                               & event rate $[\mathrm{yr}^{-1}]$ & $[\mathrm{yr}^{-1}]$ & event rate $[\mathrm{yr}^{-1}]$      & $[\mathrm{yr}^{-1}]$ &  \\ \hline
Ref-L100N1504                  & 2.02                & 0.01     & 2.36                  & 0.02     &  \\
Ref-L100N1504 + Delays         & 1.89                & 0.01     & 2.17                  & 0.01     &  \\
SS-L050N0752                   & 2.02                & 0.04     & 2.16                  & 0.04     & 
\end{tabular}
\caption{Estimated event rates for the different simulation models. $\sigma$ is the standard Poisson uncertainty on the expectation event rate due to the finite volume of the simulations.}\label{Tab:event_rate}
\end{table*}

For the signals detected by eLISA it will be possible to extract the physical parameters of the BH sources, such as their masses, luminosity distance, and sky locations, using a set of theoretical templates for the waveforms for each phase (i.e inspiral, merger, and ringdown phases)  \citep{Cutler1994,Flanagan:1997,Seoane:2012GWN, LIGO}. In \cref{fig:Results} we have plotted the redshift distribution, the mass function of the more massive member of the binary, ${M}_1$, and the distribution of the mass ratio, ${M}_1/{M}_2$, of the number of detected SMBH coalescences per observed year at redshift $z = 0$ (left, central and right panels respectively). In this plot we have also included a variation of the Ref-L100N1504 simulation that include a delay to the SMBH merger timescales as detailed in \cref{sec:Delays}.

From the first panel it is clear that SMBHs merging between redshift $z\sim2$ and $z\sim2$ will provide the greatest contribution to the event rate of GW signals in the \textsc{eagle} models. On the basis of the redshift distributions of detected signals it is thus not possible to discriminate between the SMBH seeding mechanisms implemented in our simulations. The ${M}_1$ mass function of the predicted event rate has a very pronounced peak at the SMBH seed mass $m_\mathrm{seed}$ ($10^5 \, \mathrm{M}_{\sun} h^{-1}$ for the \textit{`Ref-'} model and $10^4 \,  \mathrm{M}_{\sun} h^{-1}$ for the \textit{`SS-'} model). Given the logarithmic scale of the plot, the galaxy formation model implemented in \textsc{eagle} predicts that GW signals will be dominated by the coalescence of BH seeds, which is also shown in the last panel of the figure, in which the mass ratio distribution peaks for equal mass SMBH coalescences for both models. This is a remarkable result, since it implies that the physical parameters of the GW sources recovered from the eLISA data stream will provide us with a profound insight into the nature of SMBHs and the initial mass distribution of seeds. We also find that adding a delay to the SMBH merger timescales makes no significant difference in either the predicted event rate, nor to the mass function of the detected binaries. This is to be expected, since the galaxy formation model implemented in \textsc{eagle} predicts GW signals that will be dominated by the coalescence of BH seeds. These low mass SMBHs are hosted mainly by gas rich galaxies in which planet-like migration is predicted to lead to short coalescence time-scales \citep{Colpi:2011, Mayer:2013, Colpi:2014}. 
 
We use \cref{eq:event_rate} to calculate the event rate of GW signals resolved by the eLISA mission. The results are shown in \cref{Tab:event_rate}.  By propagation of error, the uncertainty on the expectation event rate due to the finite volume of the simulations is given by
\begin{equation}\label{eq:event_rate_error}
	\sigma = \sqrt{\sum^{z=\infty}_{z=0} \left(\frac{\sqrt{\bar{N}(z,k\geq 1)}}{\Delta z \Delta V_c}\frac{\Delta z}{\Delta t}\frac{\Delta V_c}{\Delta z}\frac{\Delta z}{(1+z)}\right)^2}.
\end{equation}

The actual number of detections is drawn from a Poisson distribution that depends on the duration of the mission multiplied by the expectation rate given in the table. We find that for the estimated event rate of GWs is $\sim2$ events per year for the inspiral and the merger-ringdown phases (for both the Ref-L100N1504 with and without delays, and SS-L050N0752 simulations). We estimate that in a 3 year mission the eLISA detector should be able to resolve $\sim6$ mergers and $\sim6$ inspiral signals from SMBH mergers. Even with this low event rate the information carried by each gravitational waveform would provide us with a powerful tool to constrain the SMBH seed formation mechanisms.  

\section{Discussion and Conclusions}\label{sec:con}
Using the \textsc{eagle} simulations, a state-of-the-art cosmological hydrodynamical simulation suite, we have computed the event rate of GW signals expected from SMBH mergers that should be resolved by a space-based GW detector such as the Evolved Laser Interferometer Space Antenna, eLISA. 

The \textsc{eagle} simulations use modern smoothed particle hydrodynamics and physically motivated subgrid models to capture the unresolved physics. These simulations reproduce the observed galaxy population with unprecedented fidelity, providing a powerful tool to study galaxy formation and evolution. 

A number of SMBH seed formation mechanisms have been proposed to explain the observed population of high-redshift quasars in our Universe. These mechanisms predict different initial mass functions of BHs seeds. These characteristic BH seed mass functions and the dynamical evolution that takes place during the merging process of SMBH binaries in the centres of colliding galaxies leave a unique imprint on the GW signals predicted by the models. Therefore, the information carried by the gravitational waveforms detected by a GW detector such as eLISA will provide us with a powerful tool to discriminate between different SMBH seeding models. 

Since the processes involved in the SMBH seed formation models are not resolved by the simulations, we assume that seed BHs are produced sufficiently frequently that every halo above a certain threshold mass contains a central BH seed. In order to investigate the dependence on the assumed BH seed mass we used two simulation models using BH seeds that differ by an order of magnitude in mass. For the Ref-L100N1504 model, high-mass BH seeds $(m_{\mathrm{seed}} = 1.475 \times 10^5 \mathrm{M}_{\sun})$ were placed at the centre of every halo with total mass greater than $m_{\mathrm{halo,th}} = 1.475 \times 10^{10} \mathrm{M}_{\sun}$ that did not already contain a BH. For the SS-L050N0752 model, the BH seed mass used was $m_{\mathrm{seed}} = 1.475 \times 10^4 \mathrm{M}_{\sun}$. These BH seeds then grow by accreting gas and via mergers with other BHs. In the \textsc{eagle} models, BHs residing in low-mass haloes barely grow because star formation driven outflows are efficient and able to prevent cold gas accumulating around the BH (Bower et al. in preparation). As a result, three prominent populations of characteristic SMBH binaries build up in the simulations as a result of the halo mass-central BH mass relation in \textsc{eagle}, shown in \cref{fig:BH_Halo}. The groups are: SMBH binaries that involve two BHs not much more massive than the seed mass (\textbf{L}$+$\textbf{L}); high mass ratio binaries in which one BH is massive ($>10^7 \mathrm{M}_{\sun}$) and the other is not much more massive than the seed mass (\textbf{L}$+$\textbf{H}); and the case where both BHs are massive, with masses between $10^7$ and $10^8$ $\mathrm{M}_{\sun}$ (\textbf{H}$+$\textbf{H}). We also consider a variation of the Ref-L100N1504 reference model, where a prescription for the expected delays in the BH merger timescale has been included after their host galaxies merge. The added delays are based on the gas content in the nuclear region of the resulting galaxy after the merger.

We combine the merger rates of SMBHs in the simulations with the most recent phenomenological frequency-domain gravitational waveform model for non-precessing BH binaries described in \citet{PhenomD2} (commonly referred to as ``PhenomD''). We calculated that the merger rate of SMBHs is similar in the simulation models and will produce a low event rate of GW signals, nonetheless observable by a space-based interferometer such as eLISA. We find that the predicted event rate of GWs for the inspiral and merger-ringdown phases for both the \textit{`Ref-'} and \textit{`SS-'} models is $\sim2$ events per year. Hence, in a 3 year mission the eLISA detector should be able to resolve $\sim6$ mergers and $\sim6$ inspiral signals from SMBH coalescences. Our analysis shows that these signals will be dominated by the coalescence of BH seeds (\textbf{L}$+$\textbf{L} population of binaries) merging between redshifts $z\sim2$ and $z\sim1$. Given the difference in the BH seed mass of the models, there is a characteristic shift of a decade in the observed frequency for the whole distribution of the GW signals (\cref{fig:f_shift}). 

Compared to previous studies that propose that eLISA could distinguish between BH seed formation models based on the global properties of the merger distribution \citep[i.e][]{Sesana:2007, SesanaGair:2011}, we find that eLISA could probe BH seeds down to low redshift because the GW signals from SMBH coalescences will be dominated by mergers of BHs that have not yet experienced significant growth (see \cref{fig:BH_Halo}, \ref{fig:Results}, and Bower et al. in preparation). Hence, different physical BH seeding mechanisms could be distinguished from the detected gravitational waveforms, allowing eLISA to provide us with profound insight into the origin of SMBHs and the initial mass distribution of SMBH seeds. We find that adding a delay to the SMBH merger timescales makes no significant difference in either the predicted event rate, nor to the mass function of the detected binaries (\cref{fig:Results}).

We find that \textsc{eagle} predicts GW signals that would be best detected by eLISA, but complementary observations of GW signals in different frequency windows will enable us to fully characterise the cosmic history of SMBHs \citep{Crowder:2005,Sesana:2008,SKA:2015,Moore:2015}. For instance, pulsar timing arrays will be able to detect GWs in a lower frequency window (i.e., $f_{\mathrm{obs}} < 10^{-6} [\mathrm{Hz}]$ with $h_c > 10^{-17}$, \citealt{Hellings:1983ApJ...265L..39H,Sesana:2008,Kelley:2016}) than the SMBH mergers arising in our cosmological volume (\cref{fig:freq-hc}). On the other hand, if the initial mass function of SMBH seeds extends to masses $<10^4 \mathrm{M}_{\sun}$, intermediate frequency missions (i.e. $ 10^{-3} \lessapprox f_{\mathrm{obs}}/[\mathrm{Hz}] \lessapprox 10^{1}$ with $h_c > 10^{-25}$), like the proposed Advanced Laser Interferometer Antenna (ALIA, \citealt{ALISA:2013}), the Big Bang Observer (BBO, \citealt{BBO:2006}), and the Deci-hertz Interferometer GW Observatory (DECIGO, \citealt{DECIGO:2006}), will be suitable to detect the mergers of seeds and shed light on their initial mass function. It is also important to highlight that other GW sources, such as galactic white dwarf binaries, will also contribute to the eLISA data stream.

Since the \textsc{eagle} simulations reproduce a wide set of observational properties of the galaxy population we may expect the physics of the real Universe to be reasonably well captured by the phenomenological sub-grid models implemented in the simulations. Nevertheless, the predicted GW event rate is specific to the galaxy formation and evolution model implemented in these simulations and the sub-grid models for BH seeding and growth via accretion and mergers. In particular, in the \textsc{eagle} simulations BH seeds are placed into haloes of mass $m_{\mathrm{halo,th}} = 10^{10} \, \mathrm{M}_{\sun}$, which corresponds to a very small galaxy of stellar mass $m_* \sim 10^7 \mathrm{M}_{\odot}$. From observational constraints such galaxies are thought to be the smallest galaxies to host SMBHs at low redshift \citep{Reines:2013,Seth:2014}. In the simulation, the stellar mass of galaxies in which BHs are seeded depends little on redshift and the BH mass at birth is already 1\% relative to the stellar mass. Some BH formation models suggest that BH seeds could form even more efficiently in still lower-mass galaxies at high redshift. In this case, SMBH mergers could be more common and therefore increase our predicted GW event rate. Our predicted rates are therefore conservative. Addressing this issue in more depth would require a simulation of considerably higher resolution (and yet comparable cosmological volume) coupled to a physical model of BH seed formation. Such a simulation is currently beyond the scope of cosmological simulation codes. Fortunately, since our models predict that eLISA should be very sensitive to the initial mass distribution of BH seeds, it will probe precisely these issues and directly compliment theoretical developments. Further work using the \textsc{eagle} simulations, coupled to physical models of BH seed formation could be used to predict the GW signals from SMBH mergers that could be detected by future GW detectors.

\section*{Acknowledgements}

We are grateful to all members of the Virgo Consortium and the \textsc{eagle} collaboration who have contributed to the development of the codes and simulations used here, as well as to the people who helped with the analysis. 

We are grateful to Mark Hannam and Sebastian Khan for providing the code to calculate the PhenomD waveforms.

This work was supported by the Science and Technology Facilities Council (grant number ST/F001166/1); European Research Council (grant numbers GA 267291 \enquote{Cosmiway} and GA 278594 \enquote{GasAroundGalaxies}) and by the Interuniversity Attraction Poles Programme initiated by the Belgian Science Policy Office (AP P7/08 CHARM). RAC is a Royal Society University Research Fellow.

This work used the DiRAC Data Centric system at Durham University, operated by the Institute for Computational Cosmology on behalf of the STFC DiRAC HPC Facility (\url{http://www.dirac.ac.uk}). This equipment was funded by BIS National E-infrastructure capital grant ST/K00042X/1, STFC capital grant ST/H008519/1, and STFC DiRAC Operations grant ST/K003267/1 and Durham University. DiRAC is part of the National E-Infrastructure. We acknowledge PRACE for awarding us access to the Curie machine based in France at TGCC, CEA, Bruy\`{e}res-le-Ch\^{a}tel. 

Jaime Salcido gratefully acknowledges the financial support from the Mexican Council for Science and Technology (CONACyT), Fellow No. 218259.




\bibliographystyle{mnras}
\bibliography{biblio} 

\begin{thebibliography}{}
\makeatletter
\relax
\def\mn@urlcharsother{\let\do\@makeother \do\$\do\&\do\#\do\^\do\_\do\%\do\~}
\def\mn@doi{\begingroup\mn@urlcharsother \@ifnextchar [ {\mn@doi@}
  {\mn@doi@[]}}
\def\mn@doi@[#1]#2{\def\@tempa{#1}\ifx\@tempa\@empty \href
  {http://dx.doi.org/#2} {doi:#2}\else \href {http://dx.doi.org/#2} {#1}\fi
  \endgroup}
\def\mn@eprint#1#2{\mn@eprint@#1:#2::\@nil}
\def\mn@eprint@arXiv#1{\href {http://arxiv.org/abs/#1} {{\tt arXiv:#1}}}
\def\mn@eprint@dblp#1{\href {http://dblp.uni-trier.de/rec/bibtex/#1.xml}
  {dblp:#1}}
\def\mn@eprint@#1:#2:#3:#4\@nil{\def\@tempa {#1}\def\@tempb {#2}\def\@tempc
  {#3}\ifx \@tempc \@empty \let \@tempc \@tempb \let \@tempb \@tempa \fi \ifx
  \@tempb \@empty \def\@tempb {arXiv}\fi \@ifundefined
  {mn@eprint@\@tempb}{\@tempb:\@tempc}{\expandafter \expandafter \csname
  mn@eprint@\@tempb\endcsname \expandafter{\@tempc}}}

\bibitem[\protect\citeauthoryear{{Abbott} et~al.,}{{Abbott}
  et~al.}{2016}]{LIGO}
{Abbott} B.~P.,  et~al., 2016, \mn@doi [Physical Review Letters]
  {10.1103/PhysRevLett.116.061102}, \href
  {http://adsabs.harvard.edu/abs/2016PhRvL.116f1102A} {116, 061102}

\bibitem[\protect\citeauthoryear{{Agarwal}, {Dalla Vecchia}, {Johnson},
  {Khochfar}  \& {Paardekooper}}{{Agarwal} et~al.}{2014}]{Agarwal:2014}
{Agarwal} B.,  {Dalla Vecchia} C.,  {Johnson} J.~L.,  {Khochfar} S.,
  {Paardekooper} J.-P.,  2014, \mn@doi [\mnras] {10.1093/mnras/stu1112}, \href
  {http://adsabs.harvard.edu/abs/2014MNRAS.443..648A} {443, 648}

\bibitem[\protect\citeauthoryear{{Alexander} \& {Hickox}}{{Alexander} \&
  {Hickox}}{2012}]{Alexander:2012}
{Alexander} D.~M.,  {Hickox} R.~C.,  2012, \mn@doi [\nar]
  {10.1016/j.newar.2011.11.003}, \href
  {http://adsabs.harvard.edu/abs/2012NewAR..56...93A} {56, 93}

\bibitem[\protect\citeauthoryear{{Alvarez}, {Wise}  \& {Abel}}{{Alvarez}
  et~al.}{2009}]{Alvarez:2009}
{Alvarez} M.~A.,  {Wise} J.~H.,   {Abel} T.,  2009, \mn@doi [\apjl]
  {10.1088/0004-637X/701/2/L133}, \href
  {http://adsabs.harvard.edu/abs/2009ApJ...701L.133A} {701, L133}

\bibitem[\protect\citeauthoryear{{Amaro-Seoane} et~al.,}{{Amaro-Seoane}
  et~al.}{2012}]{Seoane:2012eLISA}
{Amaro-Seoane} P.,  et~al., 2012, \mn@doi [Classical and Quantum Gravity]
  {10.1088/0264-9381/29/12/124016}, \href
  {http://adsabs.harvard.edu/abs/2012CQGra..29l4016A} {29, 124016}

\bibitem[\protect\citeauthoryear{{Amaro-Seoane} et~al.,}{{Amaro-Seoane}
  et~al.}{2013}]{Seoane:2012GWN}
{Amaro-Seoane} P.,  et~al., 2013,
  \href{http://brownbag.lisascience.org/files/2013/05/GW_Notes_Number_06.pdf}{GW
  Notes, Vol.~6}, \href {http://adsabs.harvard.edu/abs/2013GWN.....6....4A} {6,
  4}

\bibitem[\protect\citeauthoryear{{Antonini}, {Barausse}  \& {Silk}}{{Antonini}
  et~al.}{2015}]{Delays}
{Antonini} F.,  {Barausse} E.,   {Silk} J.,  2015, \mn@doi [\apj]
  {10.1088/0004-637X/812/1/72}, \href
  {http://adsabs.harvard.edu/abs/2015ApJ...812...72A} {812, 72}

\bibitem[\protect\citeauthoryear{{Bah{\'e}} et~al.,}{{Bah{\'e}}
  et~al.}{2016}]{Bahe:2016}
{Bah{\'e}} Y.~M.,  et~al., 2016, \mn@doi [\mnras] {10.1093/mnras/stv2674},
  \href {http://adsabs.harvard.edu/abs/2016MNRAS.456.1115B} {456, 1115}

\bibitem[\protect\citeauthoryear{{Begelman}, {Blandford}  \& {Rees}}{{Begelman}
  et~al.}{1980}]{Begelman:1980}
{Begelman} M.~C.,  {Blandford} R.~D.,   {Rees} M.~J.,  1980, \mn@doi [\nat]
  {10.1038/287307a0}, \href {http://adsabs.harvard.edu/abs/1980Natur.287..307B}
  {287, 307}

\bibitem[\protect\citeauthoryear{{Begelman}, {Rossi}  \& {Armitage}}{{Begelman}
  et~al.}{2008}]{Begelman:2008}
{Begelman} M.~C.,  {Rossi} E.~M.,   {Armitage} P.~J.,  2008, \mn@doi [\mnras]
  {10.1111/j.1365-2966.2008.13344.x}, \href
  {http://adsabs.harvard.edu/abs/2008MNRAS.387.1649B} {387, 1649}

\bibitem[\protect\citeauthoryear{{Bender}, {Begelman}  \& {Gair}}{{Bender}
  et~al.}{2013}]{ALISA:2013}
{Bender} P.~L.,  {Begelman} M.~C.,   {Gair} J.~R.,  2013, \mn@doi [Classical
  and Quantum Gravity] {10.1088/0264-9381/30/16/165017}, \href
  {http://adsabs.harvard.edu/abs/2013CQGra..30p5017B} {30, 165017}

\bibitem[\protect\citeauthoryear{{Booth} \& {Schaye}}{{Booth} \&
  {Schaye}}{2009}]{Booth:2009}
{Booth} C.~M.,  {Schaye} J.,  2009, \mn@doi [\mnras]
  {10.1111/j.1365-2966.2009.15043.x}, \href
  {http://adsabs.harvard.edu/abs/2009MNRAS.398...53B} {398, 53}

\bibitem[\protect\citeauthoryear{{Bower}, {Benson}, {Malbon}, {Helly}, {Frenk},
  {Baugh}, {Cole}  \& {Lacey}}{{Bower} et~al.}{2006}]{Bower:2006}
{Bower} R.~G.,  {Benson} A.~J.,  {Malbon} R.,  {Helly} J.~C.,  {Frenk} C.~S.,
  {Baugh} C.~M.,  {Cole} S.,   {Lacey} C.~G.,  2006, \mn@doi [\mnras]
  {10.1111/j.1365-2966.2006.10519.x}, \href
  {http://adsabs.harvard.edu/abs/2006MNRAS.370..645B} {370, 645}

\bibitem[\protect\citeauthoryear{{Clark}, {Glover}, {Smith}, {Greif}, {Klessen}
   \& {Bromm}}{{Clark} et~al.}{2011}]{Clark:2011}
{Clark} P.~C.,  {Glover} S.~C.~O.,  {Smith} R.~J.,  {Greif} T.~H.,  {Klessen}
  R.~S.,   {Bromm} V.,  2011, \mn@doi [Science] {10.1126/science.1198027},
  \href {http://adsabs.harvard.edu/abs/2011Sci...331.1040C} {331, 1040}

\bibitem[\protect\citeauthoryear{{Colpi}}{{Colpi}}{2014}]{Colpi:2014}
{Colpi} M.,  2014, \mn@doi [\ssr] {10.1007/s11214-014-0067-1}, \href
  {http://adsabs.harvard.edu/abs/2014SSRv..183..189C} {183, 189}

\bibitem[\protect\citeauthoryear{{Colpi} \& {Dotti}}{{Colpi} \&
  {Dotti}}{2011}]{Colpi:2011}
{Colpi} M.,  {Dotti} M.,  2011, \mn@doi [Advanced Science Letters]
  {10.1166/asl.2011.1205}, \href
  {http://adsabs.harvard.edu/abs/2011ASL.....4..181C} {4, 181}

\bibitem[\protect\citeauthoryear{Crain et~al.}{Crain et~al.}{2015}]{Crain:2015}
Crain R.~A.,  et~al., 2015, \mn@doi [MNRAS] {10.1093/mnras/stv725}, \href
  {http://adsabs.harvard.edu/abs/2015MNRAS.450.1937C} {450, 1937}

\bibitem[\protect\citeauthoryear{{Crowder} \& {Cornish}}{{Crowder} \&
  {Cornish}}{2005}]{Crowder:2005}
{Crowder} J.,  {Cornish} N.~J.,  2005, \mn@doi [\prd]
  {10.1103/PhysRevD.72.083005}, \href
  {http://adsabs.harvard.edu/abs/2005PhRvD..72h3005C} {72, 083005}

\bibitem[\protect\citeauthoryear{{Cullen} \& {Dehnen}}{{Cullen} \&
  {Dehnen}}{2010}]{Cullen:2010}
{Cullen} L.,  {Dehnen} W.,  2010, \mn@doi [\mnras]
  {10.1111/j.1365-2966.2010.17158.x}, \href
  {http://adsabs.harvard.edu/abs/2010MNRAS.408..669C} {408, 669}

\bibitem[\protect\citeauthoryear{{Cutler} \& {Flanagan}}{{Cutler} \&
  {Flanagan}}{1994}]{Cutler1994}
{Cutler} C.,  {Flanagan} {\'E}.~E.,  1994, \mn@doi [\prd]
  {10.1103/PhysRevD.49.2658}, \href
  {http://adsabs.harvard.edu/abs/1994PhRvD..49.2658C} {49, 2658}

\bibitem[\protect\citeauthoryear{{Dalla Vecchia} \& {Schaye}}{{Dalla Vecchia}
  \& {Schaye}}{2012}]{DallaVecchiaSchaye:2012}
{Dalla Vecchia} C.,  {Schaye} J.,  2012, \mn@doi [\mnras]
  {10.1111/j.1365-2966.2012.21704.x}, \href
  {http://adsabs.harvard.edu/abs/2012MNRAS.426..140D} {426, 140}

\bibitem[\protect\citeauthoryear{{Devecchi} \& {Volonteri}}{{Devecchi} \&
  {Volonteri}}{2009}]{Devecchi:2009}
{Devecchi} B.,  {Volonteri} M.,  2009, \mn@doi [\apj]
  {10.1088/0004-637X/694/1/302}, \href
  {http://adsabs.harvard.edu/abs/2009ApJ...694..302D} {694, 302}

\bibitem[\protect\citeauthoryear{{Devecchi}, {Volonteri}, {Rossi}, {Colpi}  \&
  {Portegies Zwart}}{{Devecchi} et~al.}{2012}]{DevecchiVolo:2012}
{Devecchi} B.,  {Volonteri} M.,  {Rossi} E.~M.,  {Colpi} M.,   {Portegies
  Zwart} S.,  2012, \mn@doi [\mnras] {10.1111/j.1365-2966.2012.20406.x}, \href
  {http://adsabs.harvard.edu/abs/2012MNRAS.421.1465D} {421, 1465}

\bibitem[\protect\citeauthoryear{{Di Matteo}, {Springel}  \& {Hernquist}}{{Di
  Matteo} et~al.}{2005}]{DiMatteo:2005}
{Di Matteo} T.,  {Springel} V.,   {Hernquist} L.,  2005, \mn@doi [\nat]
  {10.1038/nature03335}, \href
  {http://adsabs.harvard.edu/abs/2005Natur.433..604D} {433, 604}

\bibitem[\protect\citeauthoryear{{Di Matteo}, {Colberg}, {Springel},
  {Hernquist}  \& {Sijacki}}{{Di Matteo} et~al.}{2008}]{DiMatteo:2008}
{Di Matteo} T.,  {Colberg} J.,  {Springel} V.,  {Hernquist} L.,   {Sijacki} D.,
   2008, \mn@doi [\apj] {10.1086/524921}, \href
  {http://adsabs.harvard.edu/abs/2008ApJ...676...33D} {676, 33}

\bibitem[\protect\citeauthoryear{{Durier} \& {Dalla Vecchia}}{{Durier} \&
  {Dalla Vecchia}}{2012}]{Durier:2012}
{Durier} F.,  {Dalla Vecchia} C.,  2012, \mn@doi [\mnras]
  {10.1111/j.1365-2966.2011.19712.x}, \href
  {http://adsabs.harvard.edu/abs/2012MNRAS.419..465D} {419, 465}

\bibitem[\protect\citeauthoryear{{Enoki}, {Inoue}, {Nagashima}  \&
  {Sugiyama}}{{Enoki} et~al.}{2004}]{Enoki:2004}
{Enoki} M.,  {Inoue} K.~T.,  {Nagashima} M.,   {Sugiyama} N.,  2004, \mn@doi
  [\apj] {10.1086/424475}, \href
  {http://adsabs.harvard.edu/abs/2004ApJ...615...19E} {615, 19}

\bibitem[\protect\citeauthoryear{{Escala}, {Larson}, {Coppi}  \&
  {Mardones}}{{Escala} et~al.}{2005}]{Escala:2005}
{Escala} A.,  {Larson} R.~B.,  {Coppi} P.~S.,   {Mardones} D.,  2005, \mn@doi
  [\apj] {10.1086/431747}, \href
  {http://adsabs.harvard.edu/abs/2005ApJ...630..152E} {630, 152}

\bibitem[\protect\citeauthoryear{{Fabian}}{{Fabian}}{2012}]{Fabian:2012}
{Fabian} A.~C.,  2012, \mn@doi [\araa] {10.1146/annurev-astro-081811-125521},
  \href {http://adsabs.harvard.edu/abs/2012ARA%26A..50..455F} {50, 455}

\bibitem[\protect\citeauthoryear{{Fan}}{{Fan}}{2006}]{Fan2006}
{Fan} X.,  2006, \mn@doi [\nar] {10.1016/j.newar.2006.06.077}, \href
  {http://adsabs.harvard.edu/abs/2006NewAR..50..665F} {50, 665}

\bibitem[\protect\citeauthoryear{{Fanidakis}, {Baugh}, {Benson}, {Bower},
  {Cole}, {Done}  \& {Frenk}}{{Fanidakis} et~al.}{2011}]{Fanidakis:2011}
{Fanidakis} N.,  {Baugh} C.~M.,  {Benson} A.~J.,  {Bower} R.~G.,  {Cole} S.,
  {Done} C.,   {Frenk} C.~S.,  2011, \mn@doi [\mnras]
  {10.1111/j.1365-2966.2010.17427.x}, \href
  {http://adsabs.harvard.edu/abs/2011MNRAS.410...53F} {410, 53}

\bibitem[\protect\citeauthoryear{{Ferrarese} \& {Merritt}}{{Ferrarese} \&
  {Merritt}}{2000}]{Ferrarese:2000}
{Ferrarese} L.,  {Merritt} D.,  2000, \mn@doi [\apjl] {10.1086/312838}, \href
  {http://adsabs.harvard.edu/abs/2000ApJ...539L...9F} {539, L9}

\bibitem[\protect\citeauthoryear{{Flanagan} \& {Hughes}}{{Flanagan} \&
  {Hughes}}{1998}]{Flanagan:1997}
{Flanagan} E.~E.,  {Hughes} S.~A.,  1998, \mn@doi [\prd]
  {10.1103/PhysRevD.57.4535}, \href
  {http://adsabs.harvard.edu/abs/1998PhRvD..57.4535F} {57, 4535}

\bibitem[\protect\citeauthoryear{{Furlong} et~al.,}{{Furlong}
  et~al.}{2015a}]{Furlong:2015-Sizes}
{Furlong} M.,  et~al., 2015a, preprint, \href
  {http://adsabs.harvard.edu/abs/2015arXiv151005645F} {} (\mn@eprint {arXiv}
  {1510.05645})

\bibitem[\protect\citeauthoryear{{Furlong} et~al.,}{{Furlong}
  et~al.}{2015b}]{Furlong:2015}
{Furlong} M.,  et~al., 2015b, \mn@doi [\mnras] {10.1093/mnras/stv852}, \href
  {http://adsabs.harvard.edu/abs/2015MNRAS.450.4486F} {450, 4486}

\bibitem[\protect\citeauthoryear{{Gair}, {Vallisneri}, {Larson}  \&
  {Baker}}{{Gair} et~al.}{2013}]{Gair:2013}
{Gair} J.~R.,  {Vallisneri} M.,  {Larson} S.~L.,   {Baker} J.~G.,  2013, Living
  Reviews in Relativity, \href
  {http://adsabs.harvard.edu/abs/2013LRR....16....7G} {16, 7}

\bibitem[\protect\citeauthoryear{{Gebhardt} et~al.,}{{Gebhardt}
  et~al.}{2000}]{Gebhardt:2000}
{Gebhardt} K.,  et~al., 2000, \mn@doi [\apjl] {10.1086/312840}, \href
  {http://adsabs.harvard.edu/abs/2000ApJ...539L..13G} {539, L13}

\bibitem[\protect\citeauthoryear{{Greif}, {Springel}, {White}, {Glover},
  {Clark}, {Smith}, {Klessen}  \& {Bromm}}{{Greif} et~al.}{2011}]{Greif:2011}
{Greif} T.~H.,  {Springel} V.,  {White} S.~D.~M.,  {Glover} S.~C.~O.,  {Clark}
  P.~C.,  {Smith} R.~J.,  {Klessen} R.~S.,   {Bromm} V.,  2011, \mn@doi [\apj]
  {10.1088/0004-637X/737/2/75}, \href
  {http://adsabs.harvard.edu/abs/2011ApJ...737...75G} {737, 75}

\bibitem[\protect\citeauthoryear{{G{\"u}ltekin} et~al.,}{{G{\"u}ltekin}
  et~al.}{2009}]{Gebhardt:2009}
{G{\"u}ltekin} K.,  et~al., 2009, \mn@doi [\apj] {10.1088/0004-637X/698/1/198},
  \href {http://adsabs.harvard.edu/abs/2009ApJ...698..198G} {698, 198}

\bibitem[\protect\citeauthoryear{{Haehnelt}}{{Haehnelt}}{1994}]{Haehnelt:1994}
{Haehnelt} M.~G.,  1994, \mnras, \href
  {http://adsabs.harvard.edu/abs/1994MNRAS.269..199H} {269, 199}

\bibitem[\protect\citeauthoryear{{Hannam}}{{Hannam}}{2014}]{Hannam2014}
{Hannam} M.,  2014, \mn@doi [General Relativity and Gravitation]
  {10.1007/s10714-014-1767-2}, \href
  {http://adsabs.harvard.edu/abs/2014GReGr..46.1767H} {46, 1767}

\bibitem[\protect\citeauthoryear{{Harry}, {Fritschel}, {Shaddock}, {Folkner}
  \& {Phinney}}{{Harry} et~al.}{2006}]{BBO:2006}
{Harry} G.~M.,  {Fritschel} P.,  {Shaddock} D.~A.,  {Folkner} W.,   {Phinney}
  E.~S.,  2006, \mn@doi [Classical and Quantum Gravity]
  {10.1088/0264-9381/23/15/008}, \href
  {http://adsabs.harvard.edu/abs/2006CQGra..23.4887H} {23, 4887}

\bibitem[\protect\citeauthoryear{{Hellings} \& {Downs}}{{Hellings} \&
  {Downs}}{1983}]{Hellings:1983ApJ...265L..39H}
{Hellings} R.~W.,  {Downs} G.~S.,  1983, \mn@doi [\apjl] {10.1086/183954},
  \href {http://adsabs.harvard.edu/abs/1983ApJ...265L..39H} {265, L39}

\bibitem[\protect\citeauthoryear{{Hoffman} \& {Loeb}}{{Hoffman} \&
  {Loeb}}{2007}]{Hoffman:2007}
{Hoffman} L.,  {Loeb} A.,  2007, \mn@doi [\mnras]
  {10.1111/j.1365-2966.2007.11694.x}, \href
  {http://adsabs.harvard.edu/abs/2007MNRAS.377..957H} {377, 957}

\bibitem[\protect\citeauthoryear{{Hopkins}}{{Hopkins}}{2013}]{Hopkins:2013}
{Hopkins} P.~F.,  2013, \mn@doi [\mnras] {10.1093/mnras/sts210}, \href
  {http://adsabs.harvard.edu/abs/2013MNRAS.428.2840H} {428, 2840}

\bibitem[\protect\citeauthoryear{{Inayoshi}, {Haiman}  \&
  {Ostriker}}{{Inayoshi} et~al.}{2016}]{Inayoshi:2016}
{Inayoshi} K.,  {Haiman} Z.,   {Ostriker} J.~P.,  2016, \mn@doi [\mnras]
  {10.1093/mnras/stw836}, \href
  {http://adsabs.harvard.edu/abs/2016MNRAS.459.3738I} {459, 3738}

\bibitem[\protect\citeauthoryear{{Janssen} et~al.,}{{Janssen}
  et~al.}{2015}]{SKA:2015}
{Janssen} G.,  et~al., 2015, Advancing Astrophysics with the Square Kilometre
  Array (AASKA14), \href {http://adsabs.harvard.edu/abs/2015aska.confE..37J}
  {p.~37}

\bibitem[\protect\citeauthoryear{{Johnson}, {Greif}  \& {Bromm}}{{Johnson}
  et~al.}{2008}]{Johnson:2008}
{Johnson} J.~L.,  {Greif} T.~H.,   {Bromm} V.,  2008, \mn@doi [\mnras]
  {10.1111/j.1365-2966.2008.13381.x}, \href
  {http://adsabs.harvard.edu/abs/2008MNRAS.388...26J} {388, 26}

\bibitem[\protect\citeauthoryear{{Katz}, {Sijacki}  \& {Haehnelt}}{{Katz}
  et~al.}{2015}]{Katz:2015}
{Katz} H.,  {Sijacki} D.,   {Haehnelt} M.~G.,  2015, \mn@doi [\mnras]
  {10.1093/mnras/stv1048}, \href
  {http://adsabs.harvard.edu/abs/2015MNRAS.451.2352K} {451, 2352}

\bibitem[\protect\citeauthoryear{{Kauffmann} \& {Haehnelt}}{{Kauffmann} \&
  {Haehnelt}}{2000}]{Kauffmann:2000}
{Kauffmann} G.,  {Haehnelt} M.,  2000, \mn@doi [\mnras]
  {10.1046/j.1365-8711.2000.03077.x}, \href
  {http://adsabs.harvard.edu/abs/2000MNRAS.311..576K} {311, 576}

\bibitem[\protect\citeauthoryear{{Kawamura} et~al.,}{{Kawamura}
  et~al.}{2006}]{DECIGO:2006}
{Kawamura} S.,  et~al., 2006, \mn@doi [Classical and Quantum Gravity]
  {10.1088/0264-9381/23/8/S17}, \href
  {http://adsabs.harvard.edu/abs/2006CQGra..23S.125K} {23, S125}

\bibitem[\protect\citeauthoryear{{Kelley}, {Blecha}  \& {Hernquist}}{{Kelley}
  et~al.}{2016}]{Kelley:2016}
{Kelley} L.~Z.,  {Blecha} L.,   {Hernquist} L.,  2016, preprint, \href
  {http://adsabs.harvard.edu/abs/2016arXiv160601900K} {} (\mn@eprint {arXiv}
  {1606.01900})

\bibitem[\protect\citeauthoryear{{Khan}, {Husa}, {Hannam}, {Ohme},
  {P{\"u}rrer}, {Forteza}  \& {Boh{\'e}}}{{Khan} et~al.}{2016}]{PhenomD2}
{Khan} S.,  {Husa} S.,  {Hannam} M.,  {Ohme} F.,  {P{\"u}rrer} M.,  {Forteza}
  X.~J.,   {Boh{\'e}} A.,  2016, \mn@doi [\prd] {10.1103/PhysRevD.93.044007},
  \href {http://adsabs.harvard.edu/abs/2016PhRvD..93d4007K} {93, 044007}

\bibitem[\protect\citeauthoryear{{King}}{{King}}{2003}]{King:2003}
{King} A.,  2003, \mn@doi [\apjl] {10.1086/379143}, \href
  {http://adsabs.harvard.edu/abs/2003ApJ...596L..27K} {596, L27}

\bibitem[\protect\citeauthoryear{{Klein} et~al.,}{{Klein}
  et~al.}{2016}]{Klein:2016}
{Klein} A.,  et~al., 2016, \mn@doi [\prd] {10.1103/PhysRevD.93.024003}, \href
  {http://adsabs.harvard.edu/abs/2016PhRvD..93b4003K} {93, 024003}

\bibitem[\protect\citeauthoryear{{Kormendy} \& {Ho}}{{Kormendy} \&
  {Ho}}{2013}]{Kormendy:2013}
{Kormendy} J.,  {Ho} L.~C.,  2013, \mn@doi [\araa]
  {10.1146/annurev-astro-082708-101811}, \href
  {http://adsabs.harvard.edu/abs/2013ARA%26A..51..511K} {51, 511}

\bibitem[\protect\citeauthoryear{{Koushiappas} \& {Zentner}}{{Koushiappas} \&
  {Zentner}}{2006}]{Koushiappas:2005}
{Koushiappas} S.~M.,  {Zentner} A.~R.,  2006, \mn@doi [\apj] {10.1086/499325},
  \href {http://adsabs.harvard.edu/abs/2006ApJ...639....7K} {639, 7}

\bibitem[\protect\citeauthoryear{{Koushiappas}, {Bullock}  \&
  {Dekel}}{{Koushiappas} et~al.}{2004}]{KBD:2004}
{Koushiappas} S.~M.,  {Bullock} J.~S.,   {Dekel} A.,  2004, \mn@doi [\mnras]
  {10.1111/j.1365-2966.2004.08190.x}, \href
  {http://adsabs.harvard.edu/abs/2004MNRAS.354..292K} {354, 292}

\bibitem[\protect\citeauthoryear{{Lagos} et~al.,}{{Lagos}
  et~al.}{2015}]{Lagos:2015}
{Lagos} C.~d.~P.,  et~al., 2015, \mn@doi [\mnras] {10.1093/mnras/stv1488},
  \href {http://adsabs.harvard.edu/abs/2015MNRAS.452.3815L} {452, 3815}

\bibitem[\protect\citeauthoryear{{Loeb} \& {Rasio}}{{Loeb} \&
  {Rasio}}{1994}]{Loeb:1994}
{Loeb} A.,  {Rasio} F.~A.,  1994, \mn@doi [\apj] {10.1086/174548}, \href
  {http://adsabs.harvard.edu/abs/1994ApJ...432...52L} {432, 52}

\bibitem[\protect\citeauthoryear{{Lupi}, {Colpi}, {Devecchi}, {Galanti}  \&
  {Volonteri}}{{Lupi} et~al.}{2014}]{Lupi:2014}
{Lupi} A.,  {Colpi} M.,  {Devecchi} B.,  {Galanti} G.,   {Volonteri} M.,  2014,
  \mn@doi [\mnras] {10.1093/mnras/stu1120}, \href
  {http://adsabs.harvard.edu/abs/2014MNRAS.442.3616L} {442, 3616}

\bibitem[\protect\citeauthoryear{{Lupi}, {Haardt}, {Dotti}, {Fiacconi}, {Mayer}
   \& {Madau}}{{Lupi} et~al.}{2016}]{Lupi:2016}
{Lupi} A.,  {Haardt} F.,  {Dotti} M.,  {Fiacconi} D.,  {Mayer} L.,   {Madau}
  P.,  2016, \mn@doi [\mnras] {10.1093/mnras/stv2877}, \href
  {http://adsabs.harvard.edu/abs/2016MNRAS.456.2993L} {456, 2993}

\bibitem[\protect\citeauthoryear{{Madau} \& {Rees}}{{Madau} \&
  {Rees}}{2001}]{Madau.and.Rees:2001}
{Madau} P.,  {Rees} M.~J.,  2001, \mn@doi [\apjl] {10.1086/319848}, \href
  {http://adsabs.harvard.edu/abs/2001ApJ...551L..27M} {551, L27}

\bibitem[\protect\citeauthoryear{{Magorrian} et~al.,}{{Magorrian}
  et~al.}{1998}]{Magorrian:1998}
{Magorrian} J.,  et~al., 1998, \mn@doi [\aj] {10.1086/300353}, \href
  {http://adsabs.harvard.edu/abs/1998AJ....115.2285M} {115, 2285}

\bibitem[\protect\citeauthoryear{{Mayer}}{{Mayer}}{2013}]{Mayer:2013}
{Mayer} L.,  2013, \mn@doi [Classical and Quantum Gravity]
  {10.1088/0264-9381/30/24/244008}, \href
  {http://adsabs.harvard.edu/abs/2013CQGra..30x4008M} {30, 244008}

\bibitem[\protect\citeauthoryear{{Mayer}, {Kazantzidis}, {Madau}, {Colpi},
  {Quinn}  \& {Wadsley}}{{Mayer} et~al.}{2007}]{Mayer:2007}
{Mayer} L.,  {Kazantzidis} S.,  {Madau} P.,  {Colpi} M.,  {Quinn} T.,
  {Wadsley} J.,  2007, \mn@doi [Science] {10.1126/science.1141858}, \href
  {http://adsabs.harvard.edu/abs/2007Sci...316.1874M} {316, 1874}

\bibitem[\protect\citeauthoryear{{McAlpine} et~al.,}{{McAlpine}
  et~al.}{2016}]{McAlpine:2015-DB}
{McAlpine} S.,  et~al., 2016, \mn@doi [Astronomy and Computing]
  {10.1016/j.ascom.2016.02.004}, \href
  {http://adsabs.harvard.edu/abs/2016A%26C....15...72M} {15, 72}

\bibitem[\protect\citeauthoryear{{Micic}, {Holley-Bockelmann}, {Sigurdsson}  \&
  {Abel}}{{Micic} et~al.}{2007}]{Micic:2007}
{Micic} M.,  {Holley-Bockelmann} K.,  {Sigurdsson} S.,   {Abel} T.,  2007,
  \mn@doi [\mnras] {10.1111/j.1365-2966.2007.12162.x}, \href
  {http://adsabs.harvard.edu/abs/2007MNRAS.380.1533M} {380, 1533}

\bibitem[\protect\citeauthoryear{{Misner}, {Thorne}  \& {Wheeler}}{{Misner}
  et~al.}{1973}]{Gravitation:1973}
{Misner} C.~W.,  {Thorne} K.~S.,   {Wheeler} J.~A.,  1973, {Gravitation}

\bibitem[\protect\citeauthoryear{{Moore}, {Cole}  \& {Berry}}{{Moore}
  et~al.}{2015}]{Moore:2015}
{Moore} C.~J.,  {Cole} R.~H.,   {Berry} C.~P.~L.,  2015, \mn@doi [Classical and
  Quantum Gravity] {10.1088/0264-9381/32/1/015014}, \href
  {http://adsabs.harvard.edu/abs/2015CQGra..32a5014M} {32, 015014}

\bibitem[\protect\citeauthoryear{{Ohme}}{{Ohme}}{2012}]{Ohme:2012}
{Ohme} F.,  2012, \mn@doi [Classical and Quantum Gravity]
  {10.1088/0264-9381/29/12/124002}, \href
  {http://adsabs.harvard.edu/abs/2012CQGra..29l4002O} {29, 124002}

\bibitem[\protect\citeauthoryear{{Omukai}}{{Omukai}}{2001}]{Omukai:2001}
{Omukai} K.,  2001, \mn@doi [\apj] {10.1086/318296}, \href
  {http://adsabs.harvard.edu/abs/2001ApJ...546..635O} {546, 635}

\bibitem[\protect\citeauthoryear{{Planck Collaboration} et~al.,}{{Planck
  Collaboration} et~al.}{2014}]{Planck}
{Planck Collaboration} et~al., 2014, \mn@doi [\aap]
  {10.1051/0004-6361/201321591}, \href
  {http://adsabs.harvard.edu/abs/2014A%26A...571A..16P} {571, A16}

\bibitem[\protect\citeauthoryear{{Price}}{{Price}}{2008}]{Price:2008}
{Price} D.~J.,  2008, \mn@doi [Journal of Computational Physics]
  {10.1016/j.jcp.2008.08.011}, \href
  {http://adsabs.harvard.edu/abs/2008JCoPh.22710040P} {227, 10040}

\bibitem[\protect\citeauthoryear{{Quinlan}}{{Quinlan}}{1996}]{Quinlan:1996}
{Quinlan} G.~D.,  1996, \mn@doi [\na] {10.1016/S1384-1076(96)00003-6}, \href
  {http://adsabs.harvard.edu/abs/1996NewA....1...35Q} {1, 35}

\bibitem[\protect\citeauthoryear{{Rahmati}, {Schaye}, {Bower}, {Crain},
  {Furlong}, {Schaller}  \& {Theuns}}{{Rahmati} et~al.}{2015}]{Rahmati:2015}
{Rahmati} A.,  {Schaye} J.,  {Bower} R.~G.,  {Crain} R.~A.,  {Furlong} M.,
  {Schaller} M.,   {Theuns} T.,  2015, \mn@doi [\mnras]
  {10.1093/mnras/stv1414}, \href
  {http://adsabs.harvard.edu/abs/2015MNRAS.452.2034R} {452, 2034}

\bibitem[\protect\citeauthoryear{{Rahmati}, {Schaye}, {Crain}, {Oppenheimer},
  {Schaller}  \& {Theuns}}{{Rahmati} et~al.}{2016}]{Rahmati:2015b}
{Rahmati} A.,  {Schaye} J.,  {Crain} R.~A.,  {Oppenheimer} B.~D.,  {Schaller}
  M.,   {Theuns} T.,  2016, \mn@doi [\mnras] {10.1093/mnras/stw453}, \href
  {http://adsabs.harvard.edu/abs/2016MNRAS.459..310R} {459, 310}

\bibitem[\protect\citeauthoryear{{Regan} \& {Haehnelt}}{{Regan} \&
  {Haehnelt}}{2009}]{Regan:2009}
{Regan} J.~A.,  {Haehnelt} M.~G.,  2009, \mn@doi [\mnras]
  {10.1111/j.1365-2966.2009.14579.x}, \href
  {http://adsabs.harvard.edu/abs/2009MNRAS.396..343R} {396, 343}

\bibitem[\protect\citeauthoryear{{Regan}, {Johansson}  \& {Wise}}{{Regan}
  et~al.}{2016}]{Regan:2015}
{Regan} J.~A.,  {Johansson} P.~H.,   {Wise} J.~H.,  2016, \mn@doi [\mnras]
  {10.1093/mnras/stw899}, \href
  {http://adsabs.harvard.edu/abs/2016MNRAS.459.3377R} {459, 3377}

\bibitem[\protect\citeauthoryear{{Reines}, {Greene}  \& {Geha}}{{Reines}
  et~al.}{2013}]{Reines:2013}
{Reines} A.~E.,  {Greene} J.~E.,   {Geha} M.,  2013, \mn@doi [\apj]
  {10.1088/0004-637X/775/2/116}, \href
  {http://adsabs.harvard.edu/abs/2013ApJ...775..116R} {775, 116}

\bibitem[\protect\citeauthoryear{{Rosas-Guevara} et~al.,}{{Rosas-Guevara}
  et~al.}{2015}]{Rosas-Guevara:2015}
{Rosas-Guevara} Y.~M.,  et~al., 2015, \mn@doi [\mnras] {10.1093/mnras/stv2056},
  \href {http://adsabs.harvard.edu/abs/2015MNRAS.454.1038R} {454, 1038}

\bibitem[\protect\citeauthoryear{{Rosas-Guevara}, {Bower}, {Schaye},
  {McAlpine}, {Dalla-Vecchia}, {Frenk}, {Schaller}  \&
  {Theuns}}{{Rosas-Guevara} et~al.}{2016}]{Rosas-Guevara:2016}
{Rosas-Guevara} Y.,  {Bower} R.~G.,  {Schaye} J.,  {McAlpine} S.,
  {Dalla-Vecchia} C.,  {Frenk} C.~S.,  {Schaller} M.,   {Theuns} T.,  2016,
  preprint, \href {http://adsabs.harvard.edu/abs/2016arXiv160400020R} {}
  (\mn@eprint {arXiv} {1604.00020})

\bibitem[\protect\citeauthoryear{{Schaller} et~al.,}{{Schaller}
  et~al.}{2015a}]{Schaller:2015}
{Schaller} M.,  et~al., 2015a, \mn@doi [\mnras] {10.1093/mnras/stv1067}, \href
  {http://adsabs.harvard.edu/abs/2015MNRAS.451.1247S} {451, 1247}

\bibitem[\protect\citeauthoryear{{Schaller}, {Dalla Vecchia}, {Schaye},
  {Bower}, {Theuns}, {Crain}, {Furlong}  \& {McCarthy}}{{Schaller}
  et~al.}{2015b}]{Schaller:2015SPH}
{Schaller} M.,  {Dalla Vecchia} C.,  {Schaye} J.,  {Bower} R.~G.,  {Theuns} T.,
   {Crain} R.~A.,  {Furlong} M.,   {McCarthy} I.~G.,  2015b, \mn@doi [\mnras]
  {10.1093/mnras/stv2169}, \href
  {http://adsabs.harvard.edu/abs/2015MNRAS.454.2277S} {454, 2277}

\bibitem[\protect\citeauthoryear{{Schaye}}{{Schaye}}{2004}]{Schaye:2004}
{Schaye} J.,  2004, \mn@doi [\apj] {10.1086/421232}, \href
  {http://adsabs.harvard.edu/abs/2004ApJ...609..667S} {609, 667}

\bibitem[\protect\citeauthoryear{{Schaye} \& {Dalla Vecchia}}{{Schaye} \&
  {Dalla Vecchia}}{2008}]{SchayeDallaVecchia:2008}
{Schaye} J.,  {Dalla Vecchia} C.,  2008, \mn@doi [\mnras]
  {10.1111/j.1365-2966.2007.12639.x}, \href
  {http://adsabs.harvard.edu/abs/2008MNRAS.383.1210S} {383, 1210}

\bibitem[\protect\citeauthoryear{{Schaye} et~al.,}{{Schaye}
  et~al.}{2015}]{Schaye:2015}
{Schaye} J.,  et~al., 2015, \mn@doi [\mnras] {10.1093/mnras/stu2058}, \href
  {http://adsabs.harvard.edu/abs/2015MNRAS.446..521S} {446, 521}

\bibitem[\protect\citeauthoryear{{Sesana}, {Volonteri}  \& {Haardt}}{{Sesana}
  et~al.}{2007}]{Sesana:2007}
{Sesana} A.,  {Volonteri} M.,   {Haardt} F.,  2007, \mn@doi [\mnras]
  {10.1111/j.1365-2966.2007.11734.x}, \href
  {http://adsabs.harvard.edu/abs/2007MNRAS.377.1711S} {377, 1711}

\bibitem[\protect\citeauthoryear{{Sesana}, {Vecchio}  \& {Colacino}}{{Sesana}
  et~al.}{2008}]{Sesana:2008}
{Sesana} A.,  {Vecchio} A.,   {Colacino} C.~N.,  2008, \mn@doi [\mnras]
  {10.1111/j.1365-2966.2008.13682.x}, \href
  {http://adsabs.harvard.edu/abs/2008MNRAS.390..192S} {390, 192}

\bibitem[\protect\citeauthoryear{{Sesana}, {Vecchio}  \& {Volonteri}}{{Sesana}
  et~al.}{2009}]{Sesana:2009-SA}
{Sesana} A.,  {Vecchio} A.,   {Volonteri} M.,  2009, \mn@doi [\mnras]
  {10.1111/j.1365-2966.2009.14499.x}, \href
  {http://adsabs.harvard.edu/abs/2009MNRAS.394.2255S} {394, 2255}

\bibitem[\protect\citeauthoryear{{Sesana}, {Gair}, {Berti}  \&
  {Volonteri}}{{Sesana} et~al.}{2011}]{SesanaGair:2011}
{Sesana} A.,  {Gair} J.,  {Berti} E.,   {Volonteri} M.,  2011, \mn@doi [\prd]
  {10.1103/PhysRevD.83.044036}, \href
  {http://adsabs.harvard.edu/abs/2011PhRvD..83d4036S} {83, 044036}

\bibitem[\protect\citeauthoryear{{Seth}, {Ag{\"u}eros}, {Lee}  \&
  {Basu-Zych}}{{Seth} et~al.}{2008}]{Seth:2008}
{Seth} A.,  {Ag{\"u}eros} M.,  {Lee} D.,   {Basu-Zych} A.,  2008, \mn@doi
  [\apj] {10.1086/528955}, \href
  {http://adsabs.harvard.edu/abs/2008ApJ...678..116S} {678, 116}

\bibitem[\protect\citeauthoryear{{Seth} et~al.,}{{Seth}
  et~al.}{2014}]{Seth:2014}
{Seth} A.~C.,  et~al., 2014, \mn@doi [\nat] {10.1038/nature13762}, \href
  {http://adsabs.harvard.edu/abs/2014Natur.513..398S} {513, 398}

\bibitem[\protect\citeauthoryear{{Shapiro} \& {Teukolsky}}{{Shapiro} \&
  {Teukolsky}}{1983}]{Shapiro:1983}
{Shapiro} S.~L.,  {Teukolsky} S.~A.,  1983, {Black holes, white dwarfs, and
  neutron stars: The physics of compact objects}

\bibitem[\protect\citeauthoryear{{Springel}}{{Springel}}{2005}]{GADGET}
{Springel} V.,  2005, \mn@doi [\mnras] {10.1111/j.1365-2966.2005.09655.x},
  \href {http://adsabs.harvard.edu/abs/2005MNRAS.364.1105S} {364, 1105}

\bibitem[\protect\citeauthoryear{{Springel}, {Di Matteo}  \&
  {Hernquist}}{{Springel} et~al.}{2005}]{Springel:2005}
{Springel} V.,  {Di Matteo} T.,   {Hernquist} L.,  2005, \mn@doi [\mnras]
  {10.1111/j.1365-2966.2005.09238.x}, \href
  {http://adsabs.harvard.edu/abs/2005MNRAS.361..776S} {361, 776}

\bibitem[\protect\citeauthoryear{{Sugimura}, {Omukai}  \& {Inoue}}{{Sugimura}
  et~al.}{2014}]{Sugimura:2014}
{Sugimura} K.,  {Omukai} K.,   {Inoue} A.~K.,  2014, \mn@doi [\mnras]
  {10.1093/mnras/stu1778}, \href
  {http://adsabs.harvard.edu/abs/2014MNRAS.445..544S} {445, 544}

\bibitem[\protect\citeauthoryear{{Tamburello}, {Capelo}, {Mayer}, {Bellovary}
  \& {Wadsley}}{{Tamburello} et~al.}{2016}]{Tamburello:2016}
{Tamburello} V.,  {Capelo} P.~R.,  {Mayer} L.,  {Bellovary} J.~M.,   {Wadsley}
  J.,  2016, preprint, \href
  {http://adsabs.harvard.edu/abs/2016arXiv160300021T} {} (\mn@eprint {arXiv}
  {1603.00021})

\bibitem[\protect\citeauthoryear{Tinto}{Tinto}{1988}]{Tinto:1988}
Tinto M.,  1988, \mn@doi [American Journal of Physics]
  {http://dx.doi.org/10.1119/1.15747}, 56, 1066

\bibitem[\protect\citeauthoryear{{Trayford} et~al.,}{{Trayford}
  et~al.}{2015}]{Trayford:2015}
{Trayford} J.~W.,  et~al., 2015, \mn@doi [\mnras] {10.1093/mnras/stv1461},
  \href {http://adsabs.harvard.edu/abs/2015MNRAS.452.2879T} {452, 2879}

\bibitem[\protect\citeauthoryear{{Trayford}, {Theuns}, {Bower}, {Crain},
  {Lagos}, {Schaller}  \& {Schaye}}{{Trayford}
  et~al.}{2016}]{Trayford:2016:Green}
{Trayford} J.~W.,  {Theuns} T.,  {Bower} R.~G.,  {Crain} R.~A.,  {Lagos}
  C.~d.~P.,  {Schaller} M.,   {Schaye} J.,  2016, \mn@doi [\mnras]
  {10.1093/mnras/stw1230}, \href
  {http://adsabs.harvard.edu/abs/2016MNRAS.tmp..895T} {}

\bibitem[\protect\citeauthoryear{{Tremmel}, {Governato}, {Volonteri}  \&
  {Quinn}}{{Tremmel} et~al.}{2015}]{Tremmel:2015}
{Tremmel} M.,  {Governato} F.,  {Volonteri} M.,   {Quinn} T.~R.,  2015, \mn@doi
  [\mnras] {10.1093/mnras/stv1060}, \href
  {http://adsabs.harvard.edu/abs/2015MNRAS.451.1868T} {451, 1868}

\bibitem[\protect\citeauthoryear{{Volonteri}}{{Volonteri}}{2010}]{Volonteri:2010}
{Volonteri} M.,  2010, \mn@doi [\aapr] {10.1007/s00159-010-0029-x}, \href
  {http://adsabs.harvard.edu/abs/2010A%26ARv..18..279V} {18, 279}

\bibitem[\protect\citeauthoryear{{Volonteri} \& {Bellovary}}{{Volonteri} \&
  {Bellovary}}{2011}]{Volonteri-Bellovay:2011-evol}
{Volonteri} M.,  {Bellovary} J.,  2011, in {von Berlepsch} R.,  ed.,  Reviews
  in Modern Astronomy Vol. 23, Reviews in Modern Astronomy. p.~189,
  \mn@doi{10.1002/9783527644384.ch11}

\bibitem[\protect\citeauthoryear{{Volonteri} \& {Bellovary}}{{Volonteri} \&
  {Bellovary}}{2012}]{VolonteriBellovary:2012}
{Volonteri} M.,  {Bellovary} J.,  2012, \mn@doi [Reports on Progress in
  Physics] {10.1088/0034-4885/75/12/124901}, \href
  {http://adsabs.harvard.edu/abs/2012RPPh...75l4901V} {75, 124901}

\bibitem[\protect\citeauthoryear{{Volonteri}, {Haardt}  \& {Madau}}{{Volonteri}
  et~al.}{2003}]{VHM:2003}
{Volonteri} M.,  {Haardt} F.,   {Madau} P.,  2003, \mn@doi [\apj]
  {10.1086/344675}, \href {http://adsabs.harvard.edu/abs/2003ApJ...582..559V}
  {582, 559}

\bibitem[\protect\citeauthoryear{{Volonteri}, {Habouzit}, {Pacucci}  \&
  {Tremmel}}{{Volonteri} et~al.}{2015a}]{Volonteri:2015edd}
{Volonteri} M.,  {Habouzit} M.,  {Pacucci} F.,   {Tremmel} M.,  2015a,
  preprint, \href {http://adsabs.harvard.edu/abs/2015arXiv151102588V} {}
  (\mn@eprint {arXiv} {1511.02588})

\bibitem[\protect\citeauthoryear{{Volonteri}, {Silk}  \& {Dubus}}{{Volonteri}
  et~al.}{2015b}]{Volonteri:2015Eddington}
{Volonteri} M.,  {Silk} J.,   {Dubus} G.,  2015b, \mn@doi [\apj]
  {10.1088/0004-637X/804/2/148}, \href
  {http://adsabs.harvard.edu/abs/2015ApJ...804..148V} {804, 148}

\bibitem[\protect\citeauthoryear{Wendland}{Wendland}{1995}]{Wendland:1995}
Wendland H.,  1995, \mn@doi [Advances in Computational Mathematics]
  {10.1007/BF02123482}, 4, 389

\bibitem[\protect\citeauthoryear{{Wiersma}, {Schaye}  \& {Smith}}{{Wiersma}
  et~al.}{2009a}]{Wiersma:2009Cooling}
{Wiersma} R.~P.~C.,  {Schaye} J.,   {Smith} B.~D.,  2009a, \mn@doi [\mnras]
  {10.1111/j.1365-2966.2008.14191.x}, \href
  {http://adsabs.harvard.edu/abs/2009MNRAS.393...99W} {393, 99}

\bibitem[\protect\citeauthoryear{{Wiersma}, {Schaye}, {Theuns}, {Dalla Vecchia}
   \& {Tornatore}}{{Wiersma} et~al.}{2009b}]{Wiersma:2009Enrichment}
{Wiersma} R.~P.~C.,  {Schaye} J.,  {Theuns} T.,  {Dalla Vecchia} C.,
  {Tornatore} L.,  2009b, \mn@doi [\mnras] {10.1111/j.1365-2966.2009.15331.x},
  \href {http://adsabs.harvard.edu/abs/2009MNRAS.399..574W} {399, 574}

\bibitem[\protect\citeauthoryear{{Wise}, {Turk}  \& {Abel}}{{Wise}
  et~al.}{2008}]{Wise:2008}
{Wise} J.~H.,  {Turk} M.~J.,   {Abel} T.,  2008, \mn@doi [\apj]
  {10.1086/588209}, \href {http://adsabs.harvard.edu/abs/2008ApJ...682..745W}
  {682, 745}

\bibitem[\protect\citeauthoryear{{Wyithe} \& {Loeb}}{{Wyithe} \&
  {Loeb}}{2003}]{WyitheLoeb:2003}
{Wyithe} J.~S.~B.,  {Loeb} A.,  2003, \mn@doi [\apj] {10.1086/375187}, \href
  {http://adsabs.harvard.edu/abs/2003ApJ...590..691W} {590, 691}

\makeatother
\end{thebibliography}




\appendix
\renewcommand{\thetable}{T\arabic{table}}

\section{Parameter variations}\label{sec:App1}

\begin{figure*}
 \includegraphics[width=0.95\textwidth]{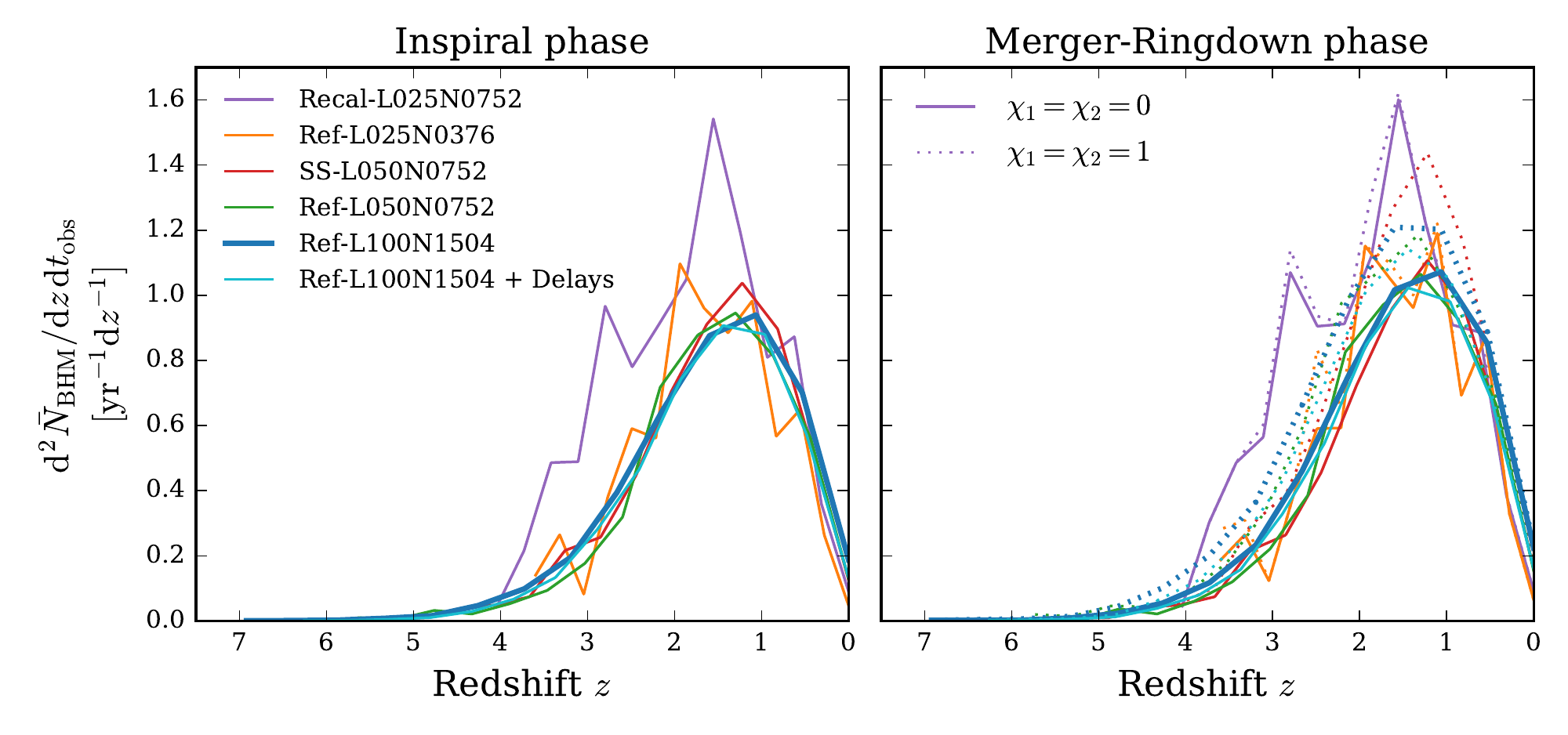}
 \caption{Number of SMBH coalescences resolved by eLISA per observed year, per unit redshift. \textit{LEFT PANEL:} inspiral phase. \textit{RIGHT PANEL:} merger-ringdown phase. Solid lines were calculated using dimensionless spin parameters $\chi_i=0$. For the dotted lines $\chi_i=1$. The distribution peaks between redshift $z\sim2$ and $z\sim1$ for both the inspiral and the merger-ringdown phases.}
 \label{fig:eventrate}
\end{figure*}

\begin{figure*}
 \includegraphics[width=0.95\textwidth]{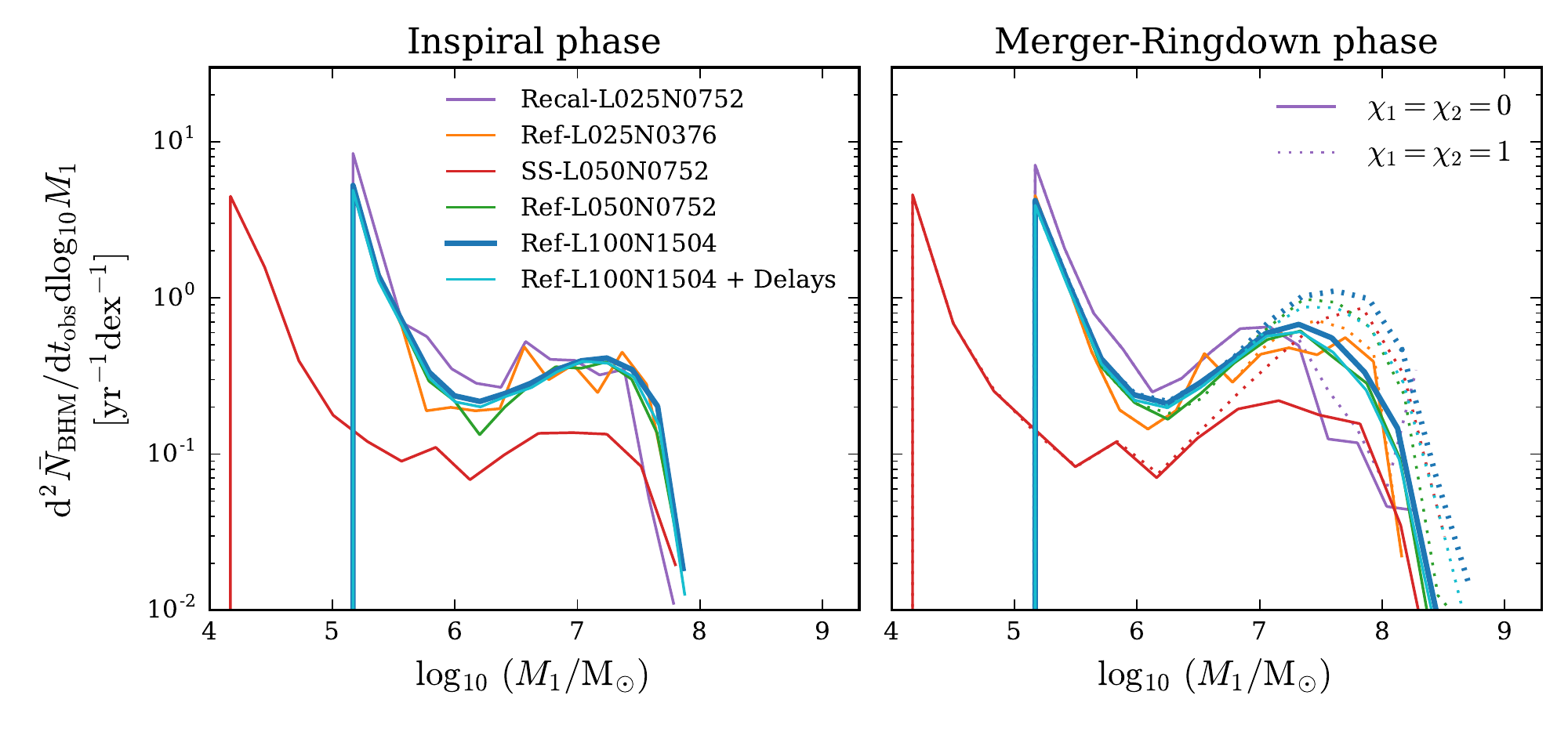}
 \caption{Distribution of mass of the more massive member of the binary, ${M}_1$, of the SMBH coalescences resolved by eLISA. \textit{LEFT PANEL:} inspiral phase. \textit{RIGHT PANEL:} merger-ringdown phase. Solid lines were calculated using  dimensionless spin parameters $\chi_i=0$. For dotted lines $\chi_i=1$. The distribution peaks at ${M}_1 \sim m_{\mathrm{seed}}$ for both the inspiral and the merger-ringdown phases.}
 \label{fig:massfunction}
\end{figure*}

\begin{figure*}
 \includegraphics[width=0.95\textwidth]{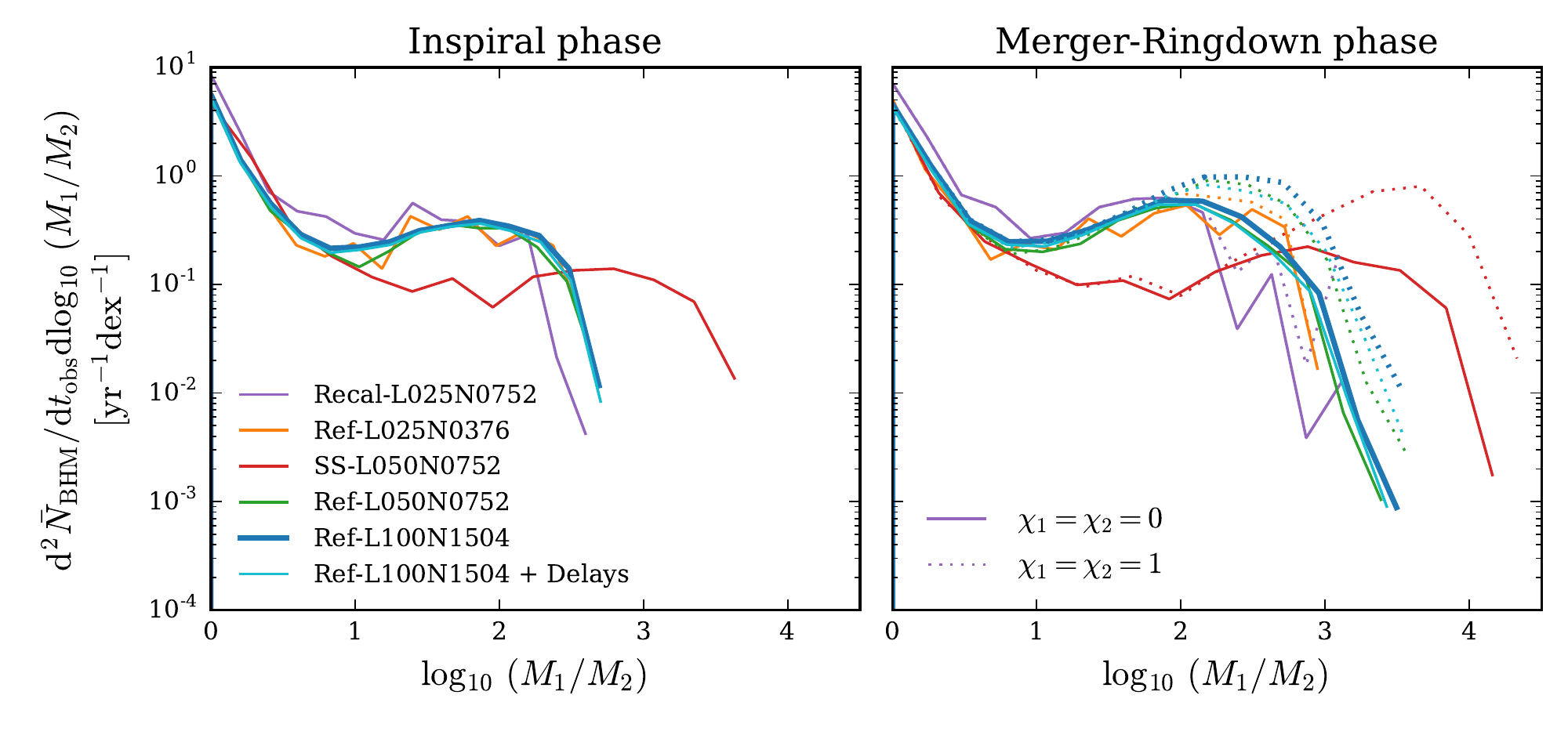}
 \caption{Distribution of the mass ratio, ${M}_1/{M}_2$, of the SMBH coalescences resolved by eLISA. \textit{LEFT PANEL:} inspiral phase. \textit{RIGHT PANEL:} merger-ringdown phase. Solid lines were calculated using dimensionless spin parameters $\chi_i=0$. For dotted lines $\chi_i=1$. The distribution peaks at equal mass SMBH binaries for both the inspiral and the merger-ringdown phase.}
 \label{fig:massratiofunction}
\end{figure*}

In this section we test our predictions against variations of the different parameters and assumptions used in our calculations. The parameters that we vary are:
\begin{itemize}
	\item Dimensionless spin parameters $(\chi_i)$
	\begin{itemize}
		\item $\chi_i=0$ for merging BHs with no spin
		\item $\chi_i=1$ for maximally spinning coalescing BHs aligned with the orbital angular momentum of the binary
	\end{itemize}
	\item Simulated volume
	\begin{itemize}
		\item Ref-L100N1504 with volume $(100 \, \mathrm{cMpc})^3$
		\item Ref-L050N0752 with volume $(50 \, \mathrm{cMpc})^3$	
		\item Ref-L025N0376 with volume $(25 \, \mathrm{cMpc})^3$
	\end{itemize} 
	\item Resolution
	\begin{itemize}
		\item  Reference model \enquote*{\textit{Ref}} (with baryonic particle mass $\mathrm{M}_\mathrm{gas} = 1.81 \times 10^6 \,\mathrm{M}_\odot$, dark matter particle mass $\mathrm{M}_\mathrm{gas} = 9.70 \times 10^6 \,\mathrm{M}_\odot$, co-moving gravitational softening $\epsilon_\mathrm{com} = 2.66 \, \mathrm{ckpc}$, and maximum proper gravitational softening $\epsilon_\mathrm{prop} = 0.70 \,\mathrm{ckpc}$) 
		\item High-resolution recalibrated model \enquote*{\textit{Recal}} (with baryonic particle mass $\mathrm{M}_\mathrm{gas} = 2.26 \times 10^5 \, \mathrm{M}_\odot$, dark matter particle mass $\mathrm{M}_\mathrm{gas} = 1.21 \times 10^6 \, \mathrm{M}_\odot$, co-moving gravitational softening $\epsilon_\mathrm{com} = 1.33 \, \mathrm{ckpc}$, and maximum proper gravitational softening $\epsilon_\mathrm{prop} = 0.35 \, \mathrm{ckpc}$). 
	\end{itemize} 
\end{itemize}

We show in \cref{fig:eventrate} the redshift distribution of the eLISA SMBH coalescence detections for both the inspiral and merger-ringdown phases. The event rates for non spinning and maximally spinning BHS ($\chi_i = 0$ and $\chi_i = 1$) are shown in the right panel. We obtain consistent results for the different \textsc{eagle} simulation models used in this study, where for both the inspiral and merger-ringdown phases the redshift distribution of the event rate peaks between redshift $z\approx2$ and $z\approx1$. There is a slight increase in the event rate when $\chi_i = 1$, which is not significant and the predicted event rate is consistent with $\sim2$.

In \cref{fig:massfunction} we show the mass function of the more massive member of the binary, ${M}_1$, for the SMBH coalescences resolved by eLISA. For the merger-ringdown phase non spinning and maximally spinning BHS ($\chi_i = 0$ and$\chi_i = 1$) are shown in the right panel. For both the inspiral and merger-ringdown phases the mass function has a very pronounced peak at ${M}_1 \sim m_{\mathrm{seed}}$. Given the logarithmic scale of the plot, the galaxy formation model implemented in \textsc{eagle} predicts that GW signals will be dominated by the coalescence of BH seeds (also shown in \cref{fig:massratiofunction}, in which the mass distribution of the mass ratio ${M}_1/{M}_2$ of the predicted event rate is dominated by equal mass SMBH coalescences).

Our event rate predictions are robust to variations of simulated volume, resolution, and parameters used to calculate the GW detections of the eLISA detector such as de dimensionless spin parameters $\chi_1$. We summarise in \cref{tab:parameter} the predicted event rates for both the inspiral and merger-ringdown phases for all the different parameter variations, simulated volumes and resolutions. Overall, the predicted event rates of GW signals resolved by the eLISA detector are $\sim2$ events per year for all the reference models, whereas the high-resolution recalibrated model Recal-L025N0752 yields $\sim3$ events per year. This does not represents a significant change in the predicted event rate.

\begin{table*}
\centering
\begin{tabular}{cclcccc}
$f_\mathrm{cut}$ & $\chi_i$ & \multicolumn{1}{c}{Simulation} & Inspiral Phase & $\sigma_\mathrm{I}$ & Merger-Ringdown Phase & $\sigma_\mathrm{M-RD}$ \\ 
\multicolumn{1}{c}{$[\mathrm{Hz}]$} & \multicolumn{1}{c}{} & \multicolumn{1}{c}{} & \multicolumn{1}{c}{event rate $[\mathrm{yr}^{-1}]$} & \multicolumn{1}{c}{$[\mathrm{yr}^{-1}]$} & \multicolumn{1}{c}{event rate $[\mathrm{yr}^{-1}]$} & \multicolumn{1}{c}{$[\mathrm{yr}^{-1}]$}\\ \hline

\multirow{12}{*}{$3\times10^{-5}$} 

& \multirow{6}{*}{0} 
& Ref-L100N1504 & 2.02 & 0.01 & 2.36  & 0.02 \\
& & Ref-L100N1504 + Delays & 1.89 & 0.01 & 2.17 & 0.01  \\
& & SS-L050N0752 & 2.02 & 0.04 & 2.16 & 0.04 \\
& & Ref-L050N0752 & 1.93 & 0.04 & 2.21 & 0.04 \\
& & Ref-L025N0376 & 1.99 & 0.11 & 2.24 & 0.11 \\
& & Recal-L025N0752 & 2.91 & 0.14 & 3.12 & 0.15\\ \cline{3-7}

& \multirow{6}{*}{1} 
& Ref-L100N1504  & 2.02 & 0.01 & 2.91 & 0.02 \\
& & Ref-L100N1504 + Delays & 1.89 & 0.01 & 2.55 & 0.02  \\
& & SS-L050N0752 & 2.02 & 0.04  & 2.75 & 0.05 \\
& & Ref-L050N0752 & 1.93 & 0.04 & 2.64 & 0.05\\
& & Ref-L025N0376 & 1.99 & 0.11 & 2.39 & 0.12\\
& & Recal-L025N0752 & 2.91 & 0.14 & 3.24 & 0.15\\ \hline

\end{tabular}
\caption{Estimated event rates for the different simulation models. $\sigma$ is the standard Poisson uncertainty on the expectation event rate due to the finite volume of the simulations.}\label{tab:parameter}
\end{table*}


\bsp	
\label{lastpage}
\end{document}